\documentclass[showpacs,amsmath,amssymb,prd,10pt,aps]{revtex4-1}
\usepackage{amssymb}
\usepackage{graphicx}
\usepackage{multirow} 
\usepackage{rotating}
\newcommand{\be}{\begin{equation}}
\newcommand{\ee}{\end{equation}}
\newcommand{\ba}{\begin{eqnarray}}
\newcommand{\ea}{\end{eqnarray}}
\newcommand{\n}{\nonumber\\}
\begin{document}
\title{Domain wall solutions to Ho\v{r}ava gravity}
\author{Carlos R. \surname{Arg\"uelles}}
\email{charly@carina.fcaglp.unlp.edu.ar}
\affiliation{Departamento de F\'\i sica - UNLP \\\it cc67, CP1900 La Plata, Argentina}
\author{Nicol\'as E. \surname{Grandi}}
\email{grandi@fisica.unlp.edu.ar}
\affiliation{\it IFLP - CONICET \,\& \,Departamento de F\'\i sica - UNLP \\\it cc67, CP1900 La Plata, Argentina}
\affiliation{Abdus Salam International Centre for Theoretical Physics, Associate Scheme \\ Strada Costiera 11, 34151, Trieste, Italy}
\begin{abstract}
We investigated purely gravitational domain wall solutions to Ho\v rava nonrelativistic theory of gravity with detailed balance in $3+1$ dimensions. We find that for arbitrary values of the running parameter $\lambda>1/3$ two branches of membrane solutions exist. For positive values of the cosmological constant, the solution represents a space that is bounded in the transverse direction, with singularities sitting at each of the boundaries. For negative values of the cosmological constant, the solution contains a single membrane sitting at the center of a space, which extends infinitely in the transverse direction approaching a Lifshitz metric. In that case there is one additional degenerate branch, for which the lapse function is undetermined.
\end{abstract}
\maketitle
\section{Introduction}
\label{sec:introduction}

The power counting renormalizable non-relativistic theory of gravity recently proposed by Ho\v rava  \cite{horava}, is a theory of gravity in which general covariance is partially abandoned in favor of renormalizability. A state of the theory is defined by a four-dimensional manifold ${\cal M}$ equip\-ped with a
three dimensional foliation ${\cal F}$, with a pseudo-Riemannian structure
defined by an Euclidean three dimensional metric in each slice of the
foliation $g_{ij}(\vec x,t)$, a shift vector $N^i(\vec x,t)$ and a
lapse function $N(\vec x,t)$. This structure can be encoded in the
ADM-decomposed metric
\be
ds^2=-N^2(\vec x,t)dt^2 + g_{ij}(\vec x,t)\left(dx^i+N^i(\vec
x,t)dt\right)\left(dx^j+N^j(\vec x,t)dt\right) \,.
\ee
The dynamics for the set $({\cal M},{\cal F},g_{ij},N_i,N)$ is
defined as being gauge invariant with respect to
foliation-preserving diffeomorfisms, and having a UV fixed point at
${\mathbf z}=3$, where the dynamical critical exponent ${\mathbf z}$ is defined as the scaling dimension of time as
compared to that of space directions $[\vec x]=-1,[t]=-{\mathbf z}$. This
choice leads to power counting renormalizability of the theory in
the UV. To the resulting action one may add relevant deformations given  by operators of lower dimensions, that lead the theory to a IR fixed
point with ${\mathbf z}=1$, in which symmetry between space and time is
restored, and thus a generally covariant theory may emerge.

In order to have control on the number of terms arising as possible
potential terms, one may impose the so-called detailed balance
condition: the potential term in the action for 3+1 dimensional
non-relativistic gravity is built from the square of the functional
derivative of a suitable action for Euclidean three-dimensional
gravity (here three-dimensional indices are contracted with the
inverse De Witt metric). Condensed matter experience on this kind of
construction tells us that the higher dimensional theory satisfying
the detailed balance condition inherits the quantum properties of the
lower dimensional one. It has to be noted that the theory is still
well defined even when detailed balance condition is broken softly, in the sense of adding relevant operators of dimension lower than that of the operators appearing at the short distance fixed
point ${\mathbf z}=3$. With such a deformation, in the UV the theory satisfies detailed balance, while in the IR the theory flows to a ${\mathbf z}=1$ fixed point.

We won't go through the above described steps in more detail, but state
the resulting action that will be
relevant to our purposes. The interested reader can  refer to the original
paper \cite{horava}. The action for non-relativistic gravity
satisfying the detailed balance condition can be written as
\small
\ba\!\!\!\!
S=\int \sqrt{g}N\left(
 \frac{2}{\kappa^2}(K_{ij}K^{ij} - \lambda K^2)
 +
 \frac{\kappa^2\mu^2(\Lambda_W R -3\Lambda_W^2)}{8(1-3\lambda)}
+
\frac{\kappa^2\mu^2 (1-4\lambda)}{32(1-3\lambda)}R^2
-
\frac{\kappa^2}{2w^4}
\left(C_{ij} -\frac{\mu w^2}{2}R_{ij}\right)
\left(C^{ij} -\frac{\mu w^2}{2}R^{ij}\right)
\right).
\label{eq:action}
\ea
\normalsize
Here $\Lambda_W$, $\kappa$, $\lambda$, $\mu$ and $w$ are arbitrary couplings, $R, R_{ij}, C_{ij}$ and $K_{ij}$ are the scalar curvature, the Ricci tensor, the Cotton-York tensor and the extrinsic curvature respectively of the three-dimensional sections of the foliation. The dynamics in the infrared is controlled by the first two terms, and then, if $\lambda=1$, general relativity is recovered. On the other hand, in the UV the third and fourth terms become dominant, and the anisotropy between space and time is explicit. 

This action is invariant under foliation preserving diffeomorphisms, namely under changes of coordinates of the form 
\ba
{x^i}'&=&{x^i}'(x^j,t)\nonumber\\
t'&=&t'(t)
\label{eq:fpd}
\ea
under which the spatial metric, shift vector and lapse function transform as 
\ba
g_{ij}'({x^{i'}},t')&=&
\partial_{i'} x^k \,
\partial_{j'} x^l g_{kl}(x^r,t)\nonumber\\
{N^{i}}'({x^{i'}},t')&=&
\partial_k x^{i'} \,
\partial_{t'} t \,
N^k(x^j,t)\nonumber\\
N'({x^{i'}},t') &=& \partial_{t'} t\, N(x^j,t)
\label{eq:traf}
\ea
Eqs. (\ref{eq:fpd}) and (\ref{eq:traf}) ensure that, if the lapse function is initially chosen to be independent of the space coordinates in a given coordinate system, it cannot be turned into a space-dependent form by a change of coordinates. In other words, the space-independence of the lapse function is a covariant statement. This implies the existence of two possible versions of Ho\v rava gravity, a ``projectable theory'' in which the lapse function is space-independent $N=N(t)$, and a ``non-projectable theory'' in which the lapse function is allowed to depend on space $N=N(x^i,t)$.

In the non-projectable case, the equations of motion obtained by varying the above action are
\be
\frac{2}{\kappa^2}(K_{ij}K^{ij} -\lambda K^2) -
\frac{\kappa^2\mu^2(\Lambda_W R
-3\Lambda_W^2)}{8(1-3\lambda)} -\frac{\kappa^2\mu^2
(1-4\lambda)}{32(1-3\lambda)}R^2 + \frac{\kappa^2}{2w^4}
Z_{ij} Z^{ij}=0\,,
\label{eq:ecdemovN}
\ee
\be
\nabla_k(K^{k\ell}-\lambda\,Kg^{k\ell})=0\,,
\label{eq:ecdemovN^i}
\ee
\be
\frac{2}{\kappa^2}E_{ij}^{(1)}-\frac{2\lambda}{\kappa^2}E_{ij}^{(2)}
+\frac{\kappa^2\mu^2\Lambda_W}{8(1-3\lambda)}E_{ij}^{(3)}
+\frac{\kappa^2\mu^2(1-4\lambda)}{32(1-3\lambda)}E_{ij}^{(4)}
-\frac{\mu\kappa^2}{4w^2}E_{ij}^{(5)}
-\frac{\kappa^2}{2w^4}E_{ij}^{(6)}=0\,,
\label{eq:ecdemovg^ij}
\ee
where
\be
Z_{ij}\equiv C_{ij} - \frac{\mu w^2}{2} R_{ij}\,,
\ee
and
\ba
E_{ij}^{(1)}&=&
N_i \nabla_k K^k{}_j + N_j\nabla_k K^k{}_i -K^k{}_i \nabla_j N_k-
   K^k{}_j\nabla_i N_k - N^k\nabla_k K_{ij}\nonumber\\
&& - 2N K_{ik} K_j{}^k
  -\frac12 N K^{k\ell} K_{k\ell}\, g_{ij} + N K K_{ij} + \dot K_{ij}
\,,\nonumber \\
E_{ij}^{(2)}&=& \frac12 NK^2 g_{ij}+ N_i \partial_j K+
N_j \partial_i K- N^k (\partial_k K)g_{ij}+  \dot K\, g_{ij}\,\,,\nonumber\\
E_{ij}^{(3)}&=&N(R_{ij}-
\frac12Rg_{ij}+\frac32\Lambda_Wg_{ij})-(
\nabla_i\nabla_j-g_{ij}\nabla_k\nabla^k)N\,,\nonumber\\
E_{ij}^{(4)}&=&NR(2R_{ij}-\frac12Rg_{ij})-
2 \big(\nabla_i\nabla_j
-g_{ij}\nabla_k\nabla^k\big)(NR)\,\,,\nonumber\\
E_{ij}^{(5)}&=&\nabla_k\big[\nabla_j(N Z^k_{~~i})
+\nabla_i(N Z^k_{~~j})\big]  -\nabla_k\nabla^k(NZ_{ij})
-\nabla_k\nabla_\ell(NZ^{k\ell})g_{ij}\,\,,\nonumber\\
E_{ij}^{(6)}&=&-\frac12NZ_{k\ell}Z^{k\ell}g_{ij}+
2NZ_{ik}Z_j^{~k}-N(Z_{ik}C_j^{~k}+Z_{jk}C_i^{~k})
+NZ_{k\ell}C^{k\ell}g_{ij}\nonumber\\
&&-\frac12\nabla_k\big[N\epsilon^{mk\ell}
(Z_{mi}R_{j\ell}+Z_{mj}R_{i\ell})\big]
+\frac12R^n{}_\ell\, \nabla_n\big[N\epsilon^{mk\ell}(Z_{mi}g_{kj}
+Z_{mj}g_{ki})\big]\nonumber\\
&&-\frac12\nabla_n\big[NZ_m^{~n}\epsilon^{mk\ell}
(g_{ki}R_{j\ell}+g_{kj}R_{i\ell})\big]
-\frac12\nabla_n\nabla^n\nabla_k\big[N\epsilon^{mk\ell}
(Z_{mi}g_{j\ell}+Z_{mj}g_{i\ell})\big]\nonumber\\
&&+\frac12\nabla_n\big[\nabla_i\nabla_k(NZ_m^{~n}\epsilon^{mk\ell})
g_{j\ell}+\nabla_j\nabla_k(NZ_m^{~n}\epsilon^{mk\ell})
g_{i\ell}\big]\nonumber\\
&&+\frac12\nabla_\ell\big[\nabla_i\nabla_k(NZ_{mj}
\epsilon^{mk\ell})+\nabla_j\nabla_k(NZ_{mi}
\epsilon^{mk\ell})\big]-\nabla_n\nabla_\ell\nabla_k
(NZ_m^{~n}\epsilon^{mk\ell})g_{ij}\,.
\ea
In the projectable case, eq.(\ref{eq:ecdemovN}) is replaced by its spatial integral.

~

Since the original proposal of \cite{horava}, there has been a growing number of research papers in the area.  Formal developments were presented in \cite{Calcagni:2009qw}-\cite{Sotiriou:2009bx}, some spherically symmetric solutions were presented in \cite{Lu:2009em}-\cite{Colgain:2009fe}, rotating solutions were studied in \cite{Ghodsi:2009zi}, string-like ans\"atze were investigated in \cite{Cho:2009fc}-\cite{Kim:2010vw}, toroidal solutions were found in \cite{Ghodsi:2009rv}, gravitational waves were studied in \cite{Nastase:2009nk} and \cite{Takahashi:2009wc}, cosmological implications were investigated in \cite{Calcagni:2009ar}-\cite{Gao:2009bx},
and interesting features of field theory in curved space and black hole physics were presented in \cite{Myung:2009dc}-\cite{Chen:2009ka}. In \cite{Blas:2009yd}-\cite{Bogdanos:2009uj} potentially harmful instabilities were pointed out, originated in the additional scalar graviton mode, that propagates in virtue of the reduced gauge symmetry. Moreover, there it was shown that the extra mode become strongly coupled at the infrared in nontrivial backgrounds. To cure these problems, a so called ``healty extension'' of the non-projectable theory was proposed, in which additional terms containing derivatives of the lapse function were included in the action, which has the effect of eliminating the instabilities \cite{Blas:2009qj}. Alternatively, a covariant theory whose partially gauge fixed version reproduces non-projectable Ho\v rava dynamics was developed, in which the extra mode was shown to be harmless \cite{Germani:2009yt}. In \cite{Bellorin:2010je}, a reinterpretation of a secondary constraint that appears when $\lambda\neq1$ in the infrared limit of the non-projectable theory, leads to a new counting of degrees of freedom in which the extra mode is not present. Finally, in \cite{Horava:2010zj} an additional $U(1)$ gauge symmetry was introduced, that kills the scalar graviton avoiding the aforementioned problems.

~

In Einstein gravity, it is easy to prove that no non-trivial solution with the symmetry of a domain wall exist in the absence of matter.  Indeed, the only solution of the equations of motion compatible with a smooth and flat domain wall ansatz is that of an AdS/Mikowski spacetime, depending on the cosmological constant. In Ho\v rava gravity on the other hand, the terms containing higher spatial derivatives could in principle play the role of a matter contribution, allowing for the existence of non-trivial domain wall solutions in vacuum. This is one of the motivations of our work.

In Einstein gravity, the knowledge of the cosmological solution corresponding to a given kind of matter can be used to obtain a domain wall solution through the so-called domain wall/cosmology correspondence \cite{Skenderis:2006fb}. Indeed, given a metric with the Friedmann-Lemaitre-Robertson-Walker form, it can be mapped to a domain wall ansatz via suitably defined Wick rotations of the coordinates. Moreover, such transformation maps the Friedman equations into the equation of motion corresponding to the domain wall, ensuring that cosmological solutions are mapped into domain wall solutions. In Ho\v rava gravity on the other hand, the situation is very different. The essential anisotropy between space and time present in the theory is an obstacle for the domain wall/cosmology correspondence to work. First, a cosmological ansatz has a single independent function that can be identified with the scale factor while, as we will see bellow in further detail, a domain wall ansatz has in principle two independent functions. Moreover, the anisotropy between space and time implies that the equations of motion for a domain wall are not mapped under Wick rotation into the Friedmann-like equations for a cosmological ansatz. In consequence, the large amount of research regarding cosmological solutions of Ho\v rava gravity \cite{Calcagni:2009ar}-\cite{Gao:2009bx} gives no clue on the form of the domain wall solutions of the theory, making the study of such solutions a subject of independent inquiry. This is a second motivation for the present paper.

In this paper we start the investigation of domain wall solutions of Ho\v rava theory. The simplest possible setup being that of a purely gravitational theory, we will not include any matter degree of freedom in our equations. In attention to the the fact that all the solutions to the aforementioned controversy about the extra scalar mode that were proposed in the literature correspond to modifications of the non-projectable theory, we will limit our investigation to that case. Since in the absence of matter, there is no possible $\mathbb{Z}_2$ symmetry, to be broken differently at each side of the wall forming ``domains'', we will sometimes use the somewhat more accurate name ``membrane'' for our solutions. 

\section{Domain wall solutions}
This paper deals with the issue of domain wall solutions of Ho\v rava theory. In the present non-relativistic context, a flat domain wall  solution is defined as a solution having translational and rotational symmetry in two dimensions, {\em i.e. } being invariant with respect to the $ISO(2)\times {\mathbb R}$ group of transformations. The symmetry of the solution is reduced with respect to that of a relativistic domain wall $ISO(2,1)$ because of the non-relativistic nature of Ho\v rava theory. An ansatz that preserves such symmetry can be easily written as
\be
ds^2=-e^{V(z)}dt^2 + e^{U(z)}\left(dx^2+dy^2\right)+ dz^2\, .
 \label{eq:anzatz}
\ee
Note that, in virtue of the reduced symmetry of the theory, there is no set of coordinates in which $U(z)=V(z)$ as it would happen in Einstein theory, and then the ansatz has two independent functions to be determined by the equations of motion. 
The variables $x,y$ can be chosen as describing a two-torus $T_2$ of volume $V_{xy}$ or a two-plane $\mathbb R^2$ (that can be considered as the infinite $V_{xy}$ limit). In what follow we will call ``lapse function'' to $e^{V(z)}$ and ``spatial volume function'' to $e^{U(z)}$.

Replacing the ansatz in the above equations (\ref{eq:ecdemovN})-(\ref{eq:ecdemovg^ij}) we get the following two independent equations of motion
\ba&&
\left(4 \Lambda_W +U'^2\right)
\left(
3 \left(4 \Lambda_W +U'^2\right)+8 U''\right)-8 (\lambda -1) U''^2=0\,,
\label{ecuaN}
\\&&\n &&48 \Lambda_W ^2
+
\left(
\left(
4 V'
-
U'
\right)U'^2
+
8\Lambda_W\left(
  U'
+
2V'
\right)\right)U'
+
8(\lambda -1)
\left(
\left( U''
-2 U'^2
-U'V'
\right)U''
-2 U'U^{(3)}\right)=0\,.
\label{ecua3}
\ea
Note that in the above equations the constants $\mu$, $\kappa$ and $w$ do not appear. This can be traced back to the action (\ref{eq:action}) or to the equations of motion (\ref{eq:ecdemovN})-(\ref{eq:ecdemovg^ij}) in which, when the extrinsic curvature $K_{ij}$ and the Cotton-York tensor $C_{ij}$ vanish, as it happens for our ansatz, the $w$ constant cancels and the product $\kappa^2\mu^2$ can be factored out.
As can be seen in the second equation, the case $\lambda=1$ is special in the fact that the total differential order of the system is reduced. As we will see, this property manifest itself in a non-analyticity of the solutions as functions of the parameter $\lambda$.

\subsection{Solutions with $\Lambda_W<0$}
\label{Lambda_neg}
\subsubsection{Solutions with $\lambda=1$}
\label{Lambda_neg_lambda_1}
We first fix our attention in the case $\lambda=1$. Solving eq.(\ref{ecuaN}) for $U(z)$ and replacing the solution into eq.(\ref{ecua3}) to get $V(z)$, we obtain the corresponding solution. It reads
\ba&&
e^{U_o(z)}= \cosh\left(\frac{3\sqrt{-\Lambda_W }}{4}  (z-z_o)\right)^{8/3}\,,
\n&&
e^{V_o(z)}=\cosh\left(\frac{3\sqrt{-\Lambda_W }}{4} (z-z_o) \right)^{2/3} \sinh\left(\frac{3\sqrt{-\Lambda_W }}{4} (z-z_o)\right)^2\,,
\label{lambda1}
\ea
where $z_o$ is a constant of integration, another two constants of integration have been reabsorbed in the definition of $t$ and of $x,y$.
We see that the solution is ${\mathbb Z}_2$ symmetric, and that its asymptotic form for $z\to\pm\infty$ is that of an AdS spacetime
\ba
&&e^{U_o(z)}\propto e^{{2\sqrt{-\Lambda_W }}|z-z_o|}\,,
\n&&
e^{V_o(z)}\propto e^{{2\sqrt{-\Lambda_W }}|z-z_o|}\,.
\ea
In order to have a physical interpretation of this solution, we evaluate some of the observable scalars of the theory. 
We start with the spacetime curvature, that reads
\be
R^{(4)}=\frac{3\Lambda_W}{4} \left(11+5\, \text{tanh}\left(\frac{3}{4} \sqrt{-\Lambda_W} (z-z_o)\right)^2\right)
\underset{{z\to\pm\infty}}\longrightarrow12\Lambda_W\,.
\ee
We see that it approaches a constant value when $z\to\pm\infty$, as may have been expected from is asymptotic AdS form. More interestingly, it shows a peak at $z=z_o$. A low energy observer, armed only with the tools of Einstein gravity, would conclude that some kind of matter with a positive energy density has to be sitting there, to partially cancel the contribution of the cosmological constant.
By observing that the total spatial volume of the $z=z_o$ slice is proportional to $V_{xy}$, 
he/she would identify the matter distribution as a ``membrane'' located at $z=z_o$. Nevertheless, from our privileged high energy point of view, we know that what we have is a purely gravitational soliton, since no additional matter has been added to Ho\v rava theory. With this at hand, we will refer to our solution as a purely gravitational membrane located at $z=z_o$.

\begin{figure}[h]
\setlength\unitlength{1mm}
\includegraphics[height=4.5cm]{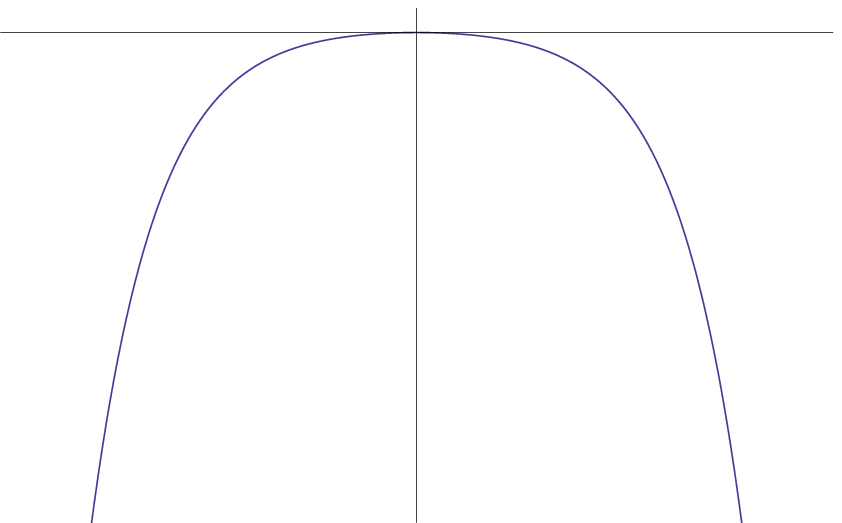}
\put(-41,46){$-e^{V_o(z)}$}
\put(2,43){$z-z_o$}\,\,\,\,\,\,\,\,\,\,\,\,
\includegraphics[height=4.5cm]{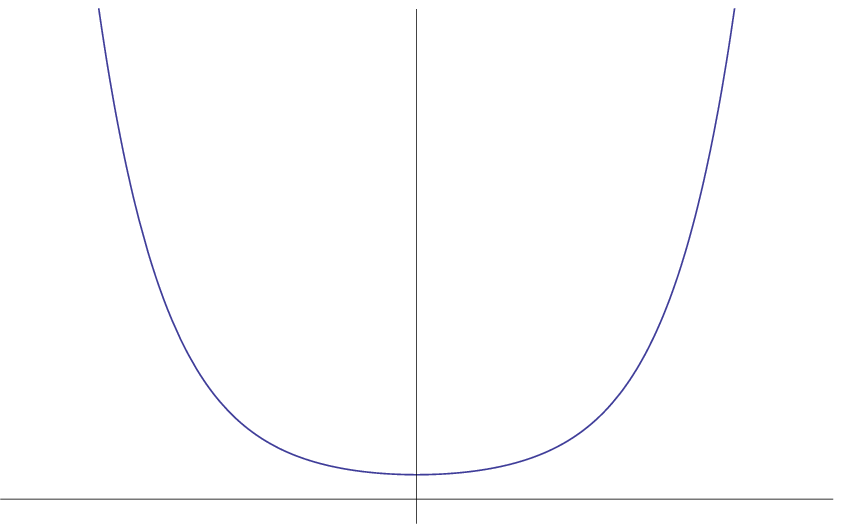}
\put(-41,46){$e^{U_o(z)}$}
\put(2,4){$z-z_o$}\\~\\~\\
\setlength\unitlength{1mm}
\includegraphics[height=4.5cm]{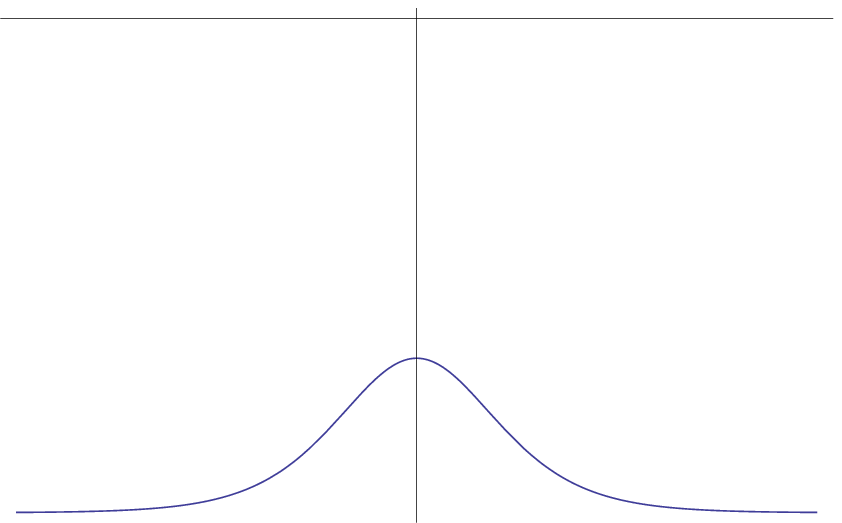}
\put(-40,46){$R^{(4)}(z)$}
\put(1,4){$z-z_o$}\,\,\,\,\,\,\,\,\,\,\,\,
\includegraphics[height=4.5cm]{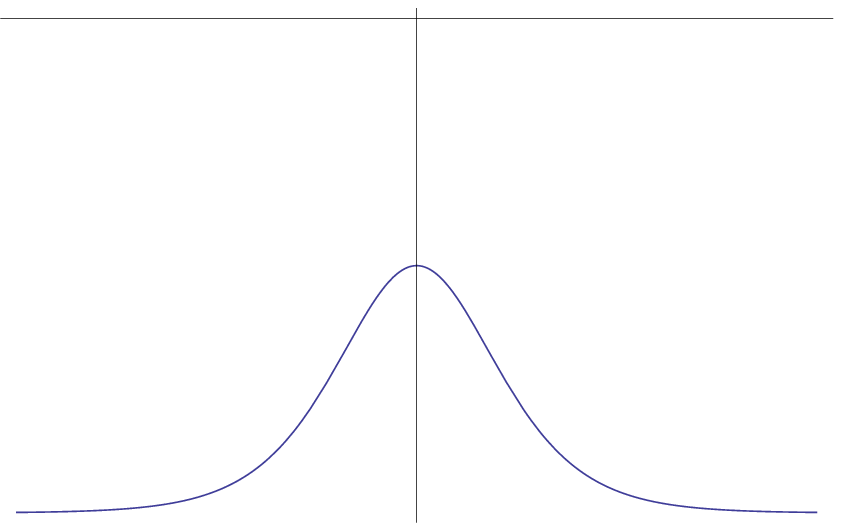}
\put(-40,46){$R^{(3)}(z)$}
\put(2,4){$z-z_o$}
\caption[FIG:A]{Plots of the lapse function $g_{00}=-e^{V_o(z)}$ (up-left) and the spatial volume function $g_{xx}=g_{yy}=e^{U_o(z)}$ (up-right), spacetime curvature $R^{(4)}(z)$ (down-left) and space curvature $R^{(3)}(z)$ (down-right), for $\Lambda_W<0$ and $\lambda=1$}.
\label{fig.LambdaNegativo.lambda1}
\end{figure}

Since Ho\v rava theory distinguishes explicitly time from space, a separated invariant of interest is the space curvature, that in our case is given by
\be
R^{(3)}=
3 \Lambda_W \left(1+\text{tanh}\left(\frac{3}{4} \sqrt{-\Lambda_W} (z-z_o)\right)^2\right)\underset{{z\to\pm\infty}}\longrightarrow6\Lambda_W\,.
\ee
Again it presents a peak at $z=z_o$, that reinforces the identification of that point as the location of the membrane. Plots of the space and spacetime curvatures have been included in Fig.\ref{fig.LambdaNegativo.lambda1}
%

~

This is all we can say about our solution from a high energy point of view. A low energy observer on the other hand, will experience a world in which the velocity of light constitutes an upper bound to the propagation of signals, and where in consequence the light-cones of the metric determine the causal structure. From that perspective he/she may wonder about the presence of horizons and/or hidden regions. To investigate such possibility, we calculate the time taken by a light signal emitted for an observer that sits at an arbitrary point $z_1$ to reach a different point $z_2$. It is given by $\Delta\tau(z_2,z_1)=e^{\frac{V_o(z_1)}2}(t(z_2)-t(z_1))$, where
\ba
t(z) &=& \int^{z}\!\!\!dz \;e^{-\frac{V_o(z)}2} 
= \n
&=&-\frac{2}{3\sqrt{-\Lambda_W}}
\ (-1)^{\frac13} B_{\!-1/\xi(z)}\!\!\left[\frac23,\frac13\right] 
\ea
where we have defined the notation 
$\xi(z)=\sinh^2\left({3\sqrt{\Lambda_W}}(z-z_o)/4\right)$, and $B_x[p,q]$ is the Euler incomplete Beta function. When $z\to z_o$, we have $\xi(z)\to 0$ and $B_{\!-1/\xi(z)}[2/3,1/3] \sim -(-1)^{\frac13}\log[\xi(z)]$ diverges. This implies that the time taken by a light signal to reach the slice at the center of the geometry $z=z_o$ is infinite, which allows us to identify such slice as a horizon. 

At this point a remark is in order: following the intuition gained on the study covariant theories, one may think that a change of variables such like $r^2=e^{U_o(z)}$ can be used to analytically extend the solution beyond the horizon, to the region $r<1$. In that region, the $t$ variable would become spacelike, while the $r$ variable would become timelike, implying that the equations of motion satisfied by the metric (\ref{ecuaN})-(\ref{ecua3}) contain third order time derivatives. But, by construction, Ho\v rava action in eq.(\ref{eq:action}) cannot give rise to equations of motion with more than two time derivatives. In consequence, in the region $r<1$ the analytically continued metric is not a solution of action (\ref{eq:action}). This implies that such an analytic continuation is not allowed in Ho\v rava theory.

To clarify the causal structure of a metric, the usual technique in relativistic theories is to isolate two of the dimensions of the metric (time and one of the space dimensions, in our case it would be $z$), and then perform a change of coordinates in order to bring the metric of the $z,t$ slices into a conformally flat form, with the new coordinates running in a compact interval. The resulting Penrose diagram is then easily drawn, by stripping the conformal factor. Indeed, if one defines new coordinates as
\ba
u_\pm&=&{\rm Sgn}(z)\,{\rm Tanh}\left( e^{ -t(z)\pm t}
\right)
\label{eq:mal}
\ea
the resulting metric reads
\be
ds^2 = \frac{e^{2  t(z)+V(z)}}{(1-u_+^2)(1-u_-^2)}\,du_+du_-+e^{U(z)}(dx^2+dy^2)
\label{eq:metrica_mal}
\ee
where now the $u_\pm$ variables are defined inside the square $[-1,1]\times[-1,1]$, in which light rays propagate along the $u_\pm$ directions. The the conformal factor of the $u_+,u_-$ slice can now be stripped, which allows for the construction of the corresponding Penrose diagram. It is given in Fig.\ref{fig:penrose1}. 

Nevertheless, at this point a very important remark is in order: in Ho\v rava theory, changes of coordinates that mix time with space do not leave the action invariant, {\em i.e.} they are not symmetries of the theory. In consequence metric (\ref{eq:metrica_mal}), obtained from (\ref{eq:anzatz}) by the illegal change of coordinates (\ref{eq:mal}), is not guaranteed to be a solution of the equations of motion. For that reason the Penrose diagram in Fig.\ref{fig:penrose1} can be considered only as an approximation valid at low energies, where general relativity is recovered. In order to have a description of the causal structure valid at any energy, the most one can do is to draw a diagram showing the light cones of the metric in the original coordinates, or in some new coordinates obtained from them via foliation preserving diffeomorphisms. We included such diagram in Fig.\ref{fig:penrose1}.
%

\subsubsection{Solutions with $\lambda\neq1$}
\label{Lambda_neg_lambda_neq_1}
Let us next assume that $\lambda\neq1$. Then
solving eq.(\ref{ecuaN}) for $U(z)$ and replacing the solution into eq.(\ref{ecua3}) to get $V(z)$ we have
\ba&&
e^{U_\pm(z)}=
\cosh\left(
\frac{\sqrt{-\Lambda_W} }{p_\pm(\lambda)}\,(z-z_o)
\right)^{2p_\pm(\lambda)}\,,
\n&&
e^{V_\pm(z)} = \left({\cosh\left(\frac{\sqrt{-\Lambda_W} }{p_\pm(\lambda)}(z-z_o)\right)^{\frac{5 p_\pm(\lambda )-4\lambda-2}{3p_\pm(\lambda)-2}} \sinh\left(\frac{\sqrt{-\Lambda_W}}{p_\pm(\lambda )}(z-z_o)\right)}\right)^2=N^2\,,
\label{solution}
\ea
with
\be
p_\pm(\lambda)=\frac{2 (\lambda -1)}{-2\pm \sqrt{6 \lambda -2}}\,,
\label{ppm}
\ee
and where again $z_o$ is a constant of integration and another two constants have been reabsorbed in the definition of $t$ and of $x,y$. Here the subindex $\pm$ indicates that we have two different branches of solutions according to the choice of sign in $p_\pm(\lambda)$. Note that for $\lambda<1/3$ no real solution exists, so in the remaining of this subsection we will focus on the region of parameters $1\neq\lambda\geq1/3$.

Both $\pm$ solutions are $\mathbb Z_2$ symmetric and centered at $z_o$. To explore the asymptotic behavior we take $z\to\pm\infty$ to have
\ba&&
e^{U_\pm(z)}\propto e^{
2\sqrt{-\Lambda_W} \,{\rm sign}(p_\pm(\lambda))\,|z-z_o|}\,,
\n&&
e^{V_\pm(z)}\propto e^{2{\mathbf z}_\pm(\lambda) \sqrt{-\Lambda_W} \,  
\,{\rm sign}(p_\pm(\lambda))
\,|z-z_o|}\,.
\ea
where
\be
{\mathbf z}_\pm(\lambda) = \frac{4(2 p_\pm(\lambda )-\lambda-1)}{p_\pm(\lambda)(3p_\pm(\lambda)-2)}\,.
\ee
Here we see that our metric corresponds to an asymptotically 
Lifshitz spacetime, similar to those studied in \cite{Kachru:2008yh}, 
 whose scaling exponent is given by ${\mathbf z}_\pm(\lambda)$. At this point, it is convenient to stress that such scaling exponent is not in principle related with the dynamical critical exponent ${\mathbf z}=3$ of Ho\v rava theory.
\begin{figure}[h]
\setlength\unitlength{1mm}
\includegraphics[height=45mm]{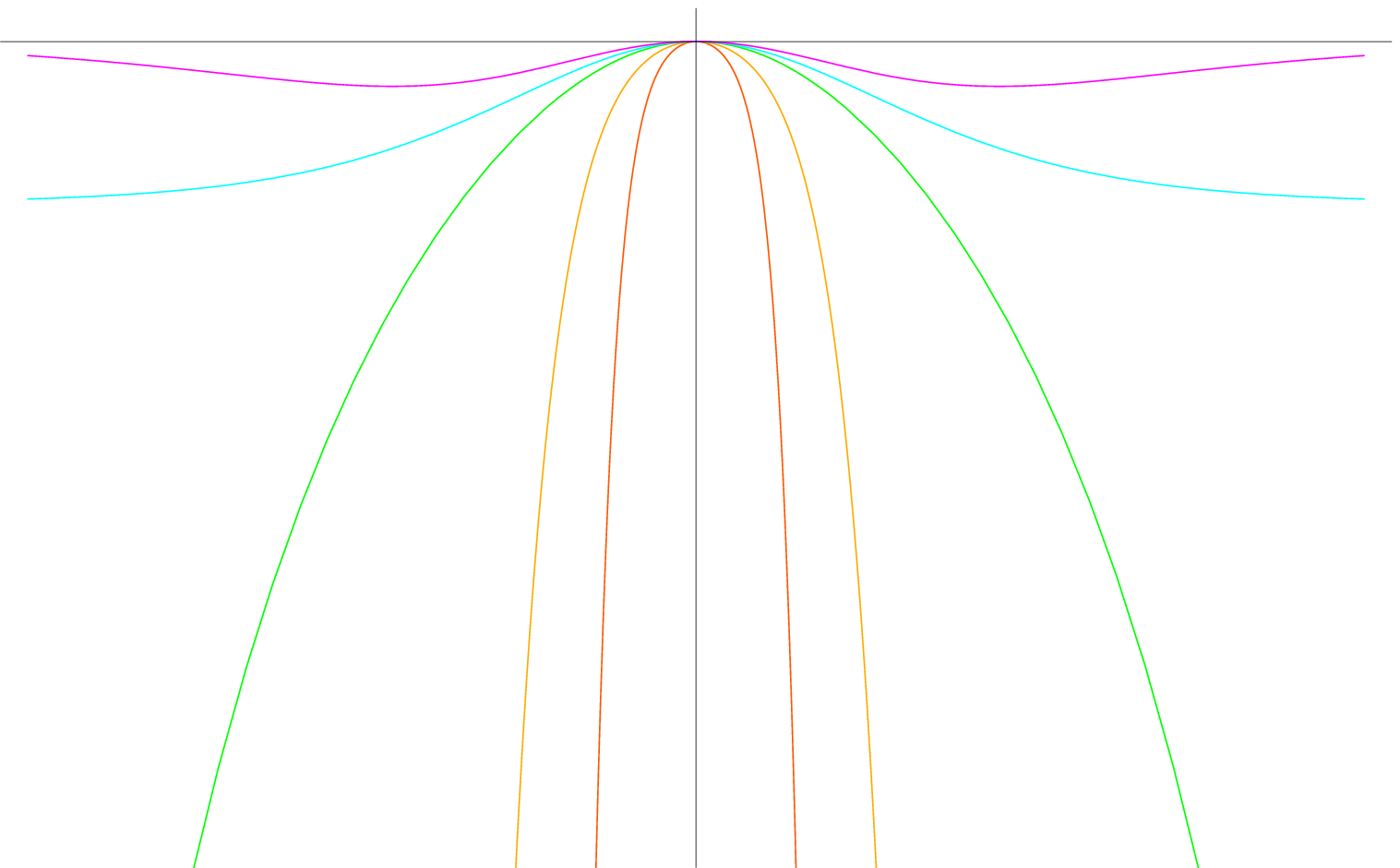}
\put(-41,46){$-e^{V_+(z)}$}
\put(2,43){$z-z_o$}\,\,\,\,\,\,\,\,\,\,\,\,
\includegraphics[height=45mm]{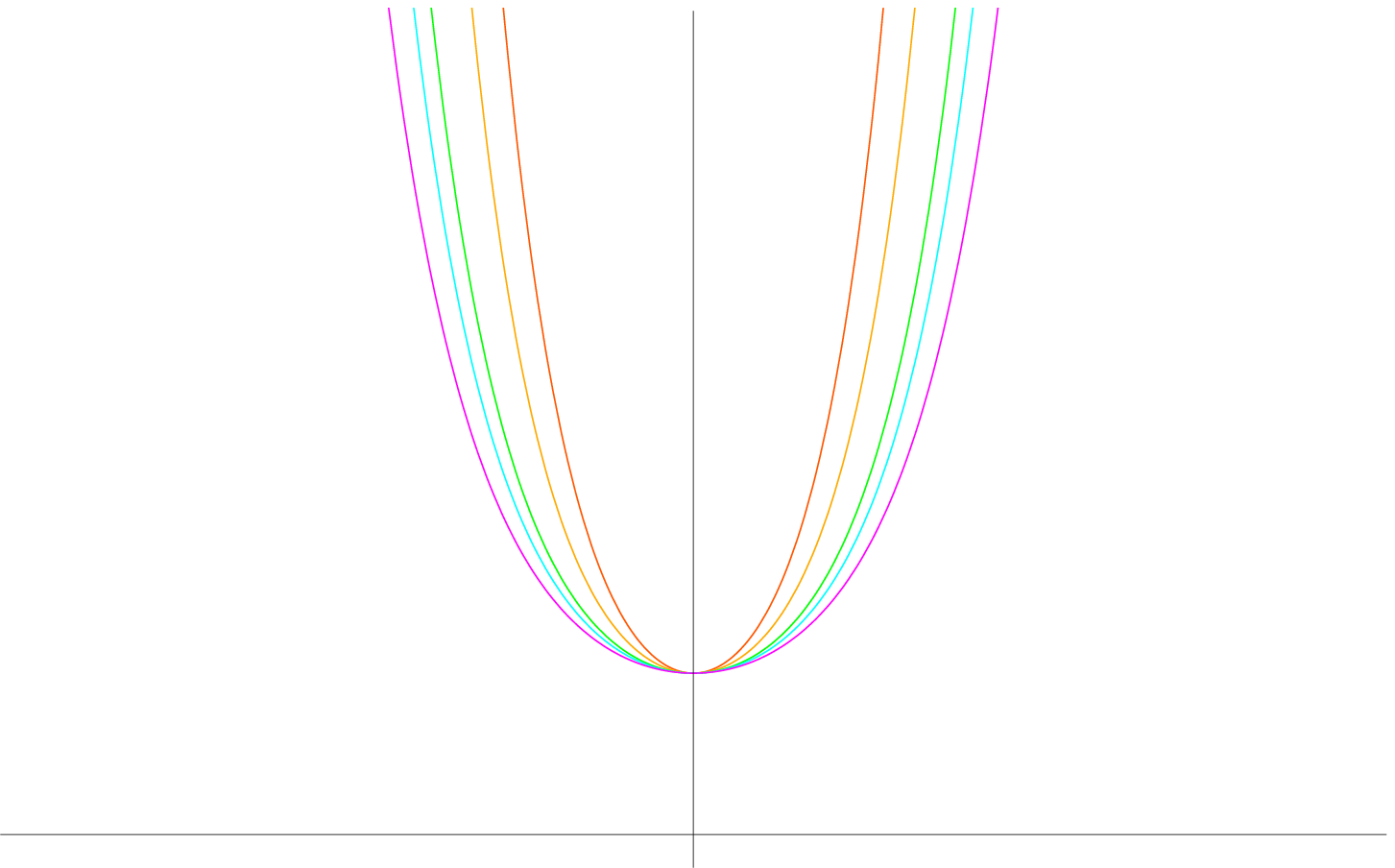}
\put(-43,46){$e^{U_+(z)}$}
\put(2,2){$z-z_o$}\\~\\~\\
\setlength\unitlength{1mm}
\includegraphics[height=45mm]{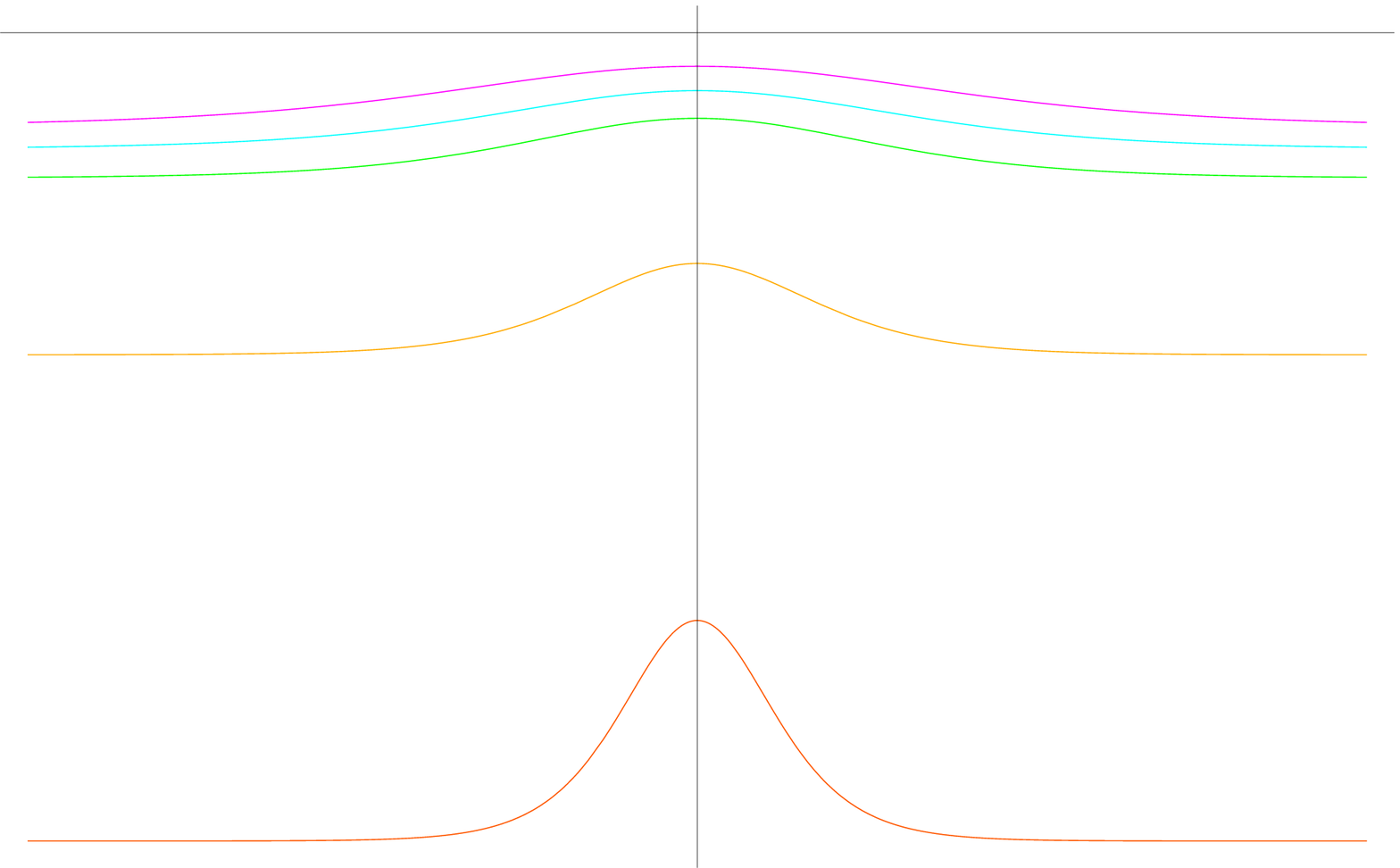}
\put(-42,46){$R^{(4)}(z)$}
\put(1,2){$z-z_o$}\,\,\,\,\,\,\,\,\,\,\,\,
\includegraphics[height=4.5cm]{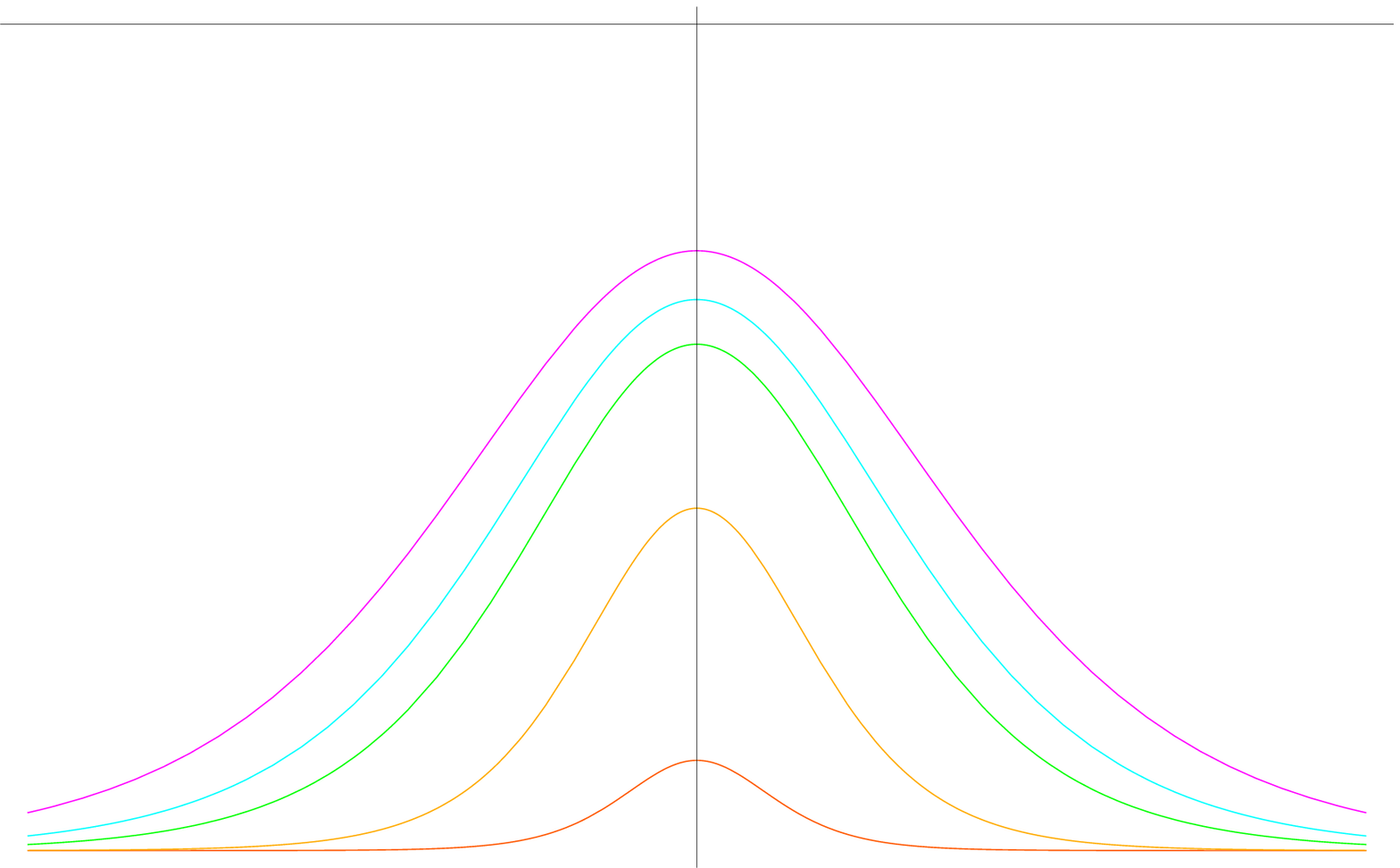}
\put(-42,46){$R^{(3)}(z)$}
\put(2,2){$z-z_o$}
\caption[FIG:A]{Plots of the lapse function $g_{00}=-e^{V_+(z)}$ (up-left) and the spatial volume function $g_{xx}=g_{yy}=e^{U_+(z)}$ (up-right), spacetime curvature $R^{(4)}(z)$ (down-left) and space curvature $R^{(3)}(z)$ (down-right) for $\Lambda_W<0$, for different values of $\lambda$. Notice the change of the asymptotic behavior of $e^{V_+(z)}$ at $\lambda=3$ in the upper left plot.}
\label{fig.LambdaNegativo.+}
\end{figure}

In the solution with the $+$ sign, $p_+(\lambda)$ is always positive, and then the spatial volume function is exponentially growing when $z\to\pm\infty$ independently of the value of the parameter $\lambda$. 
The lapse function instead grows exponentially whenever ${\mathbf z}_+(\lambda)>0$ ($\lambda<3$) and decreases exponentially for ${\mathbf z}_+(\lambda)<0$ ($\lambda>3$). In the limiting case ${\mathbf z}_+(\lambda)=0$
($\lambda=3$) the function asymptotes a constant value. Plots of the corresponding solutions for different values of $\lambda$ can be seen in Fig.\ref{fig.LambdaNegativo.+}.
\begin{figure}[h]
\setlength\unitlength{1mm}
\includegraphics[height=4.5cm]{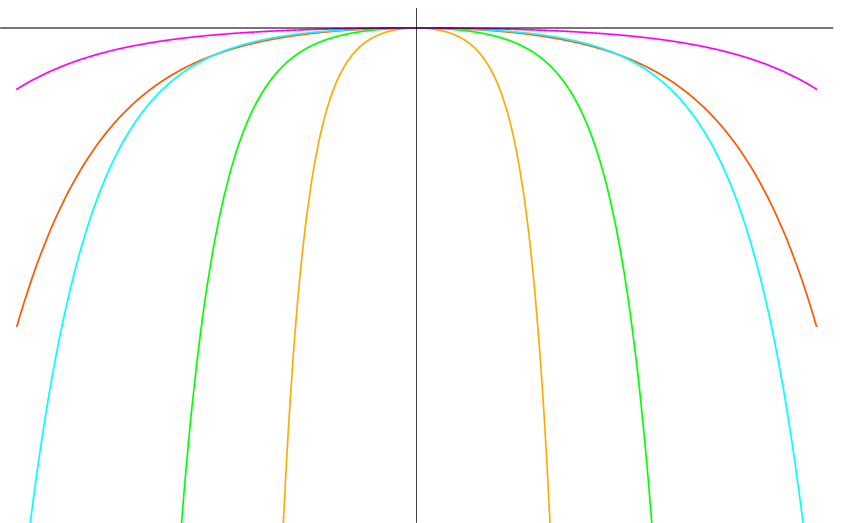}
\put(-41,46){$-e^{V_-(z)}$}
\put(2,43){$z-z_o$}\,\,\,\,\,\,\,\,\,\,\,\,
\includegraphics[height=4.5cm]{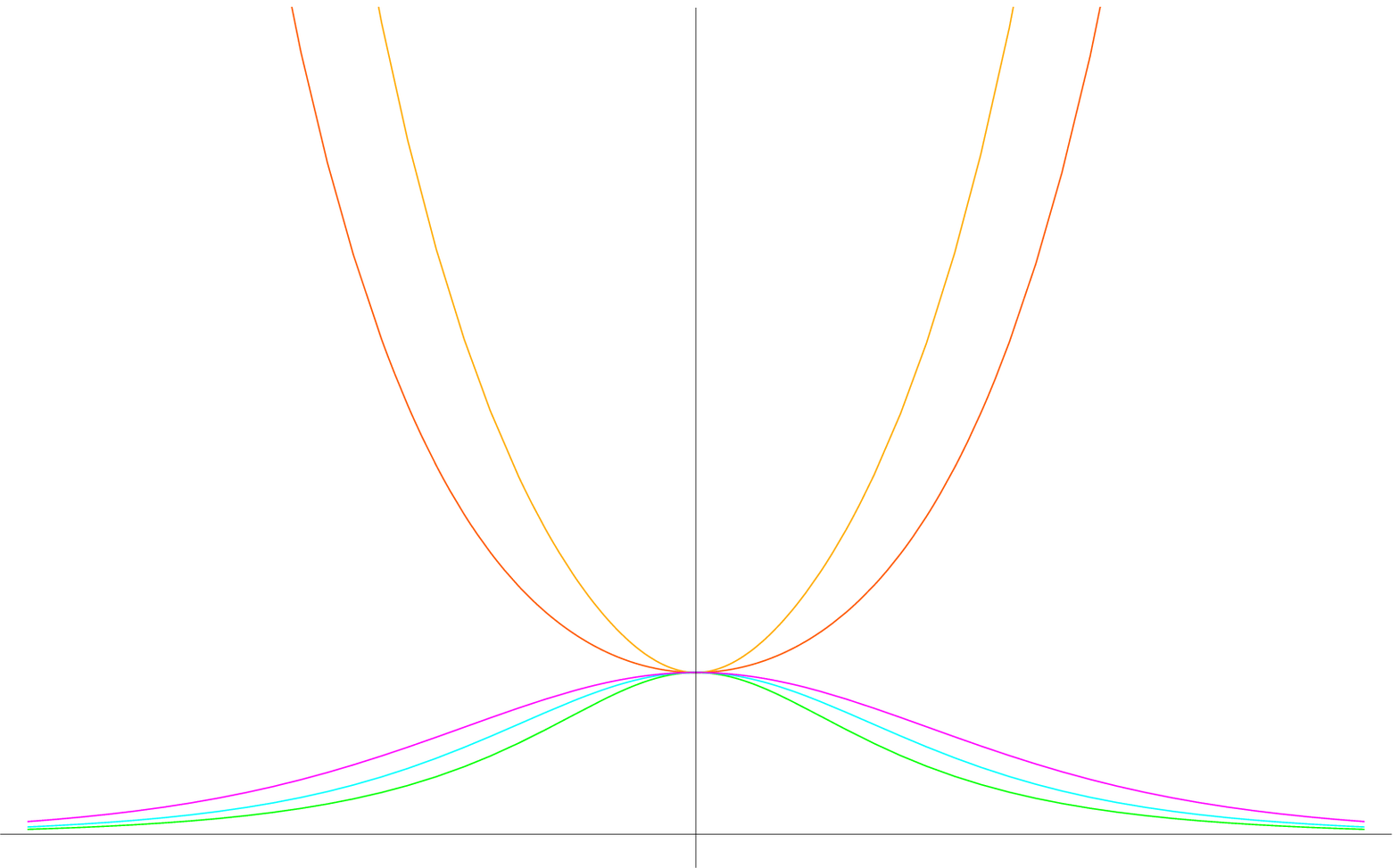}
\put(-43,46){$e^{U_-(z)}$}
\put(2,2){$z-z_o$}\\~\\~\\
\setlength\unitlength{1mm}
\includegraphics[height=4.5cm]{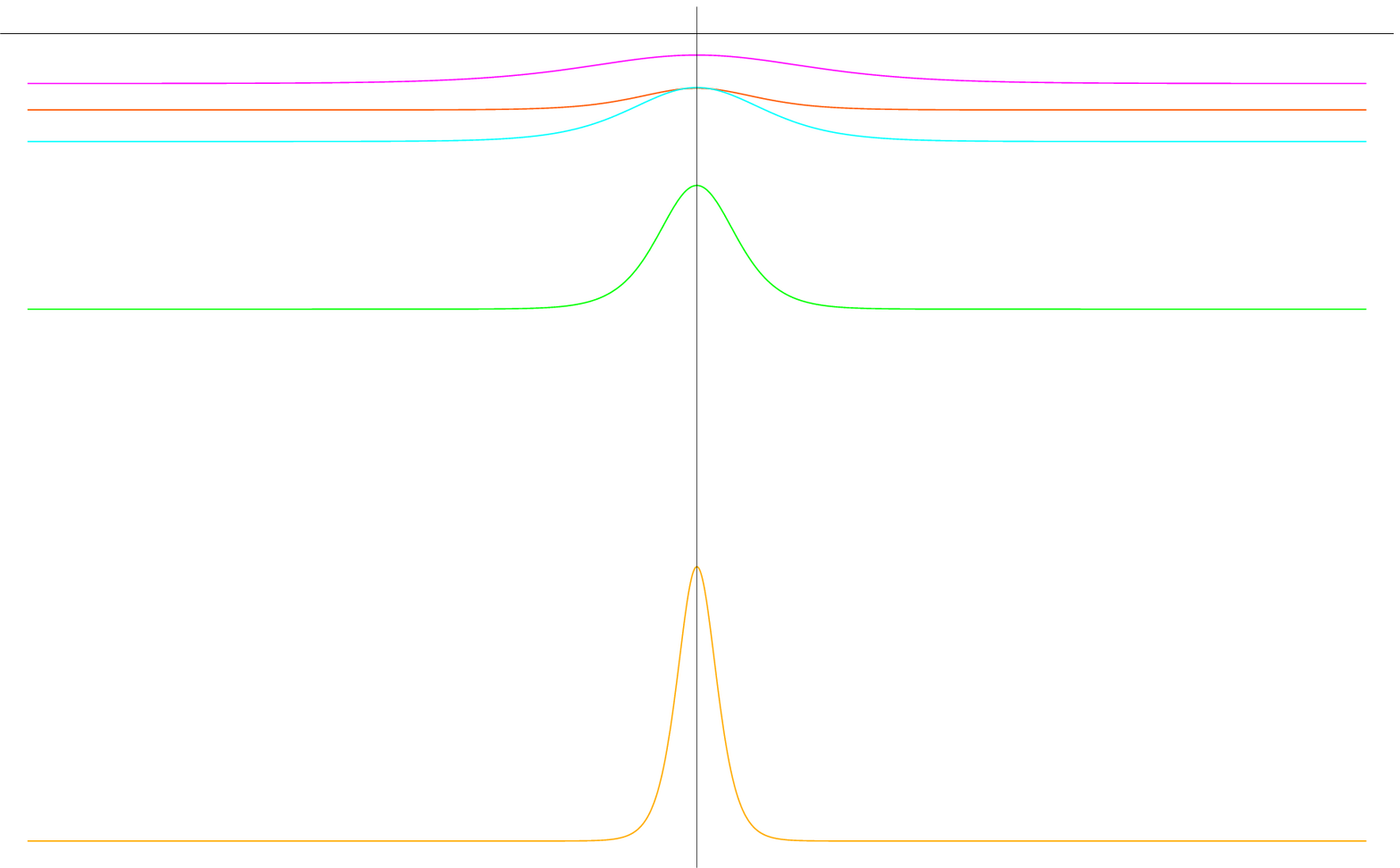}
\put(-41,46){$R^{(4)}(z)$}
\put(2,40){$z-z_o$}\,\,\,\,\,\,\,\,\,\,\,\,
\includegraphics[height=4.5cm]{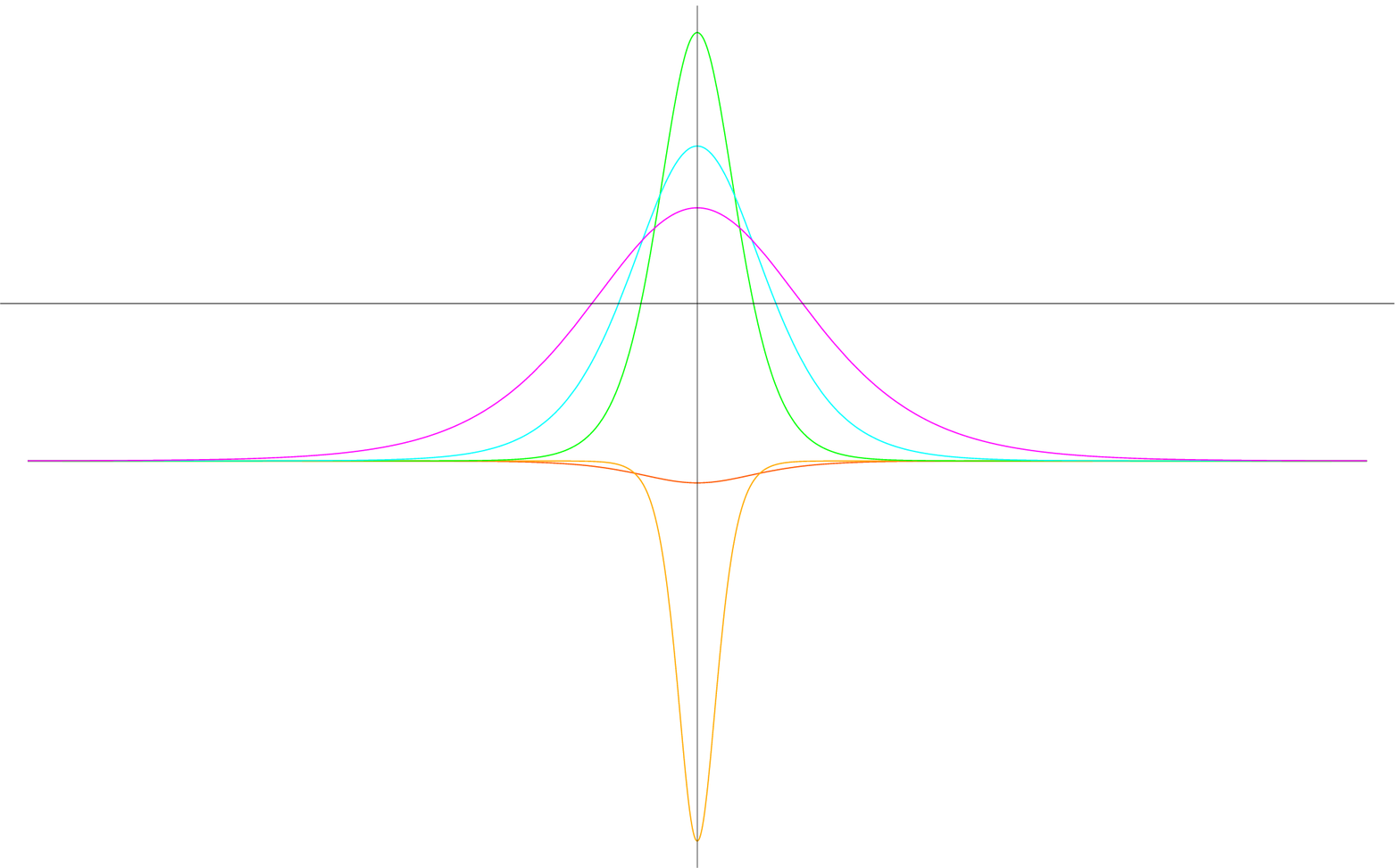}
\put(-42,46){$R^{(3)}(z)$}
\put(2,21){$z-z_o$}
\caption[FIG:A]{Plots of the lapse function $g_{00}=-e^{V_-(z)}$ (up-left) and th spatial volume function $g_{xx}=g_{yy}=e^{U_-(z)}$ (up-right), spacetime curvature $R^{(4)}(z)$ (down-left) and space curvature $R^{(3)}(z)$ (down-right) for $\Lambda_W<0$ for different values of $\lambda$. Notice the change of the asymptotic behavior of $e^{U_-(z)}$ at $\lambda=1$ in the upper right plot.}
\label{fig.LambdaNegativo.-}
\end{figure}

On the other hand, in the solution with the $-$ sign, the spatial volume function is exponentially decreasing for $z\to\pm\infty$ when $p_-(\lambda)<0$ ($\lambda>1$), and exponentially growing for $p_-(\lambda)>0$ ($\lambda<1$). Additionally, ${\mathbf z}_-(\lambda)$ is always positive, which implies that the lapse function grows exponentially when $z\to\pm\infty$ for any value of $\lambda$.  Plots of the corresponding solutions for different values of $\lambda$ can be seen in Fig.\ref{fig.LambdaNegativo.-}.

Again, to have a physical interpretation, we evaluate the spacetime curvature scalar for this solutions. It reads
\be
R^{(4)}=R^{(4)}_\infty(\lambda)+(R^{(4)}_{z_o}(\lambda)-R^{(4)}_\infty(\lambda))\, \text{sech}\left(\frac{\sqrt{-\Lambda_W} (z-z_o)}{p_\pm(\lambda)}\right)^2\,.
\ee
Its asymptotic value is consistent with a Lifshitz spacetime and it is given by
\be
R^{(4)}_\infty(\lambda)=\frac{-2\Lambda_W\left(-16 (1+\lambda )^2+p_\pm(\lambda)(48 (1+\lambda )-p_\pm(\lambda) (20-24 \lambda +3 p_\pm(\lambda)(4+9 p_\pm(\lambda))))\right)}{(2-3 p_\pm(\lambda))^2 p_\pm(\lambda)^2}\,.
\ee
As before, a peak at $z=z_o$ appears, with the value 
\be
R^{(4)}_{z_o}(\lambda)=R^{(4)}_\infty(\lambda)-
\frac{2\Lambda_W\left(8 \lambda  (1+2 \lambda )+p_\pm(\lambda)(-4 (1+3 \lambda )+p_\pm(\lambda) (14-24 \lambda +3 p_\pm(\lambda) (-8+9 p_\pm(\lambda))))
\right)}{(2-3 p_\pm(\lambda))^2 p_\pm(\lambda)^2}\,.
\ee
Such peak,
together with the observation that the spatial volume of the $z=z_o$ slice is proportional to $V_{xy}$, 
allows us to make the same interpretation as before, referring to our solution as a purely gravitational membrane sitting at $z=z_o$. This interpretation is reinforced by evaluating the space curvature scalar, that is given by
\be
R^{(3)}=
2\Lambda_W\left(3+\frac{2-3 p_\pm(\lambda)}{p_\pm(\lambda)}\,\text{sech}\left(\frac{\sqrt{-\Lambda_W} (z-{z_o})}{p_\pm(\lambda)}\right)^2\right)
\underset{z\to\infty}\longrightarrow=6\Lambda_W\,,
\ee
and that also shows a peak at $z=z_o$. 
As can be easily seen in the above equation, the space curvature at infinity is independent of $\lambda$. On the other hand its sign at $z_o$ depends on $\lambda$, for the solution with the $+$ sign it is always positive, while for the solution with the $-$ sign it is positive for $\lambda>1$ and negative otherwise.

Plots of both spacetime and space curvatures for each kind of solution $\pm$ were included in Figs. \ref{fig.LambdaNegativo.+} and \ref{fig.LambdaNegativo.-}.

~

\begin{figure}[h]
\vskip-1.8cm
\setlength\unitlength{1mm}
\includegraphics[height=10cm]{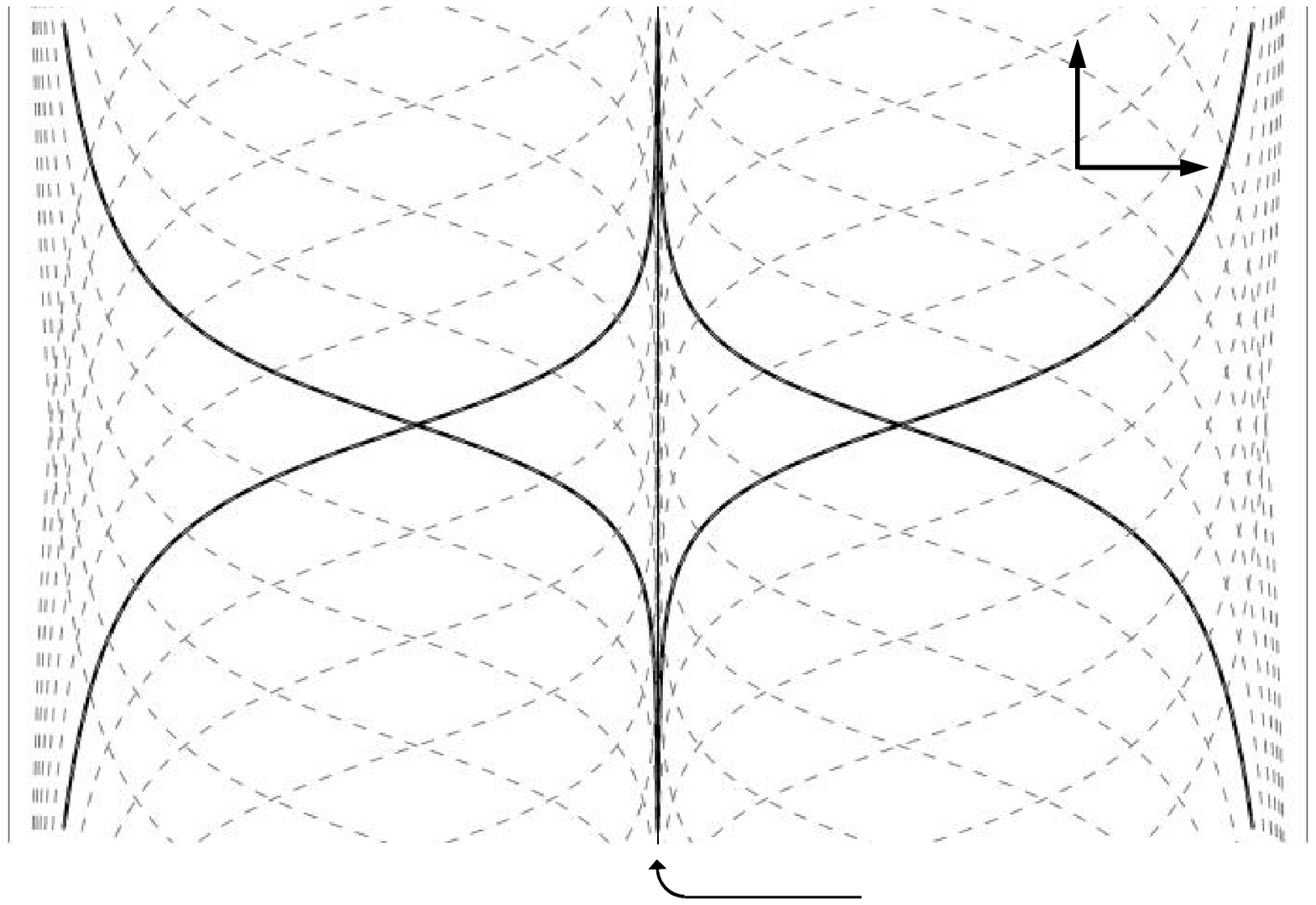}
\put(-25,48){Horizon}
\put(-18,85){$t$}
\put(-12,78){$z$}
\put(-4.5,63){{\begin{sideways}Infinity\end{sideways}}}
\put(-67.2,63){{\begin{sideways}Infinity\end{sideways}}}
\,\,\,\,\,\,\,\,\,\,\,\,
\,\,\,\,\,\,\,\,\,\,\,\,
\includegraphics[height=11cm]{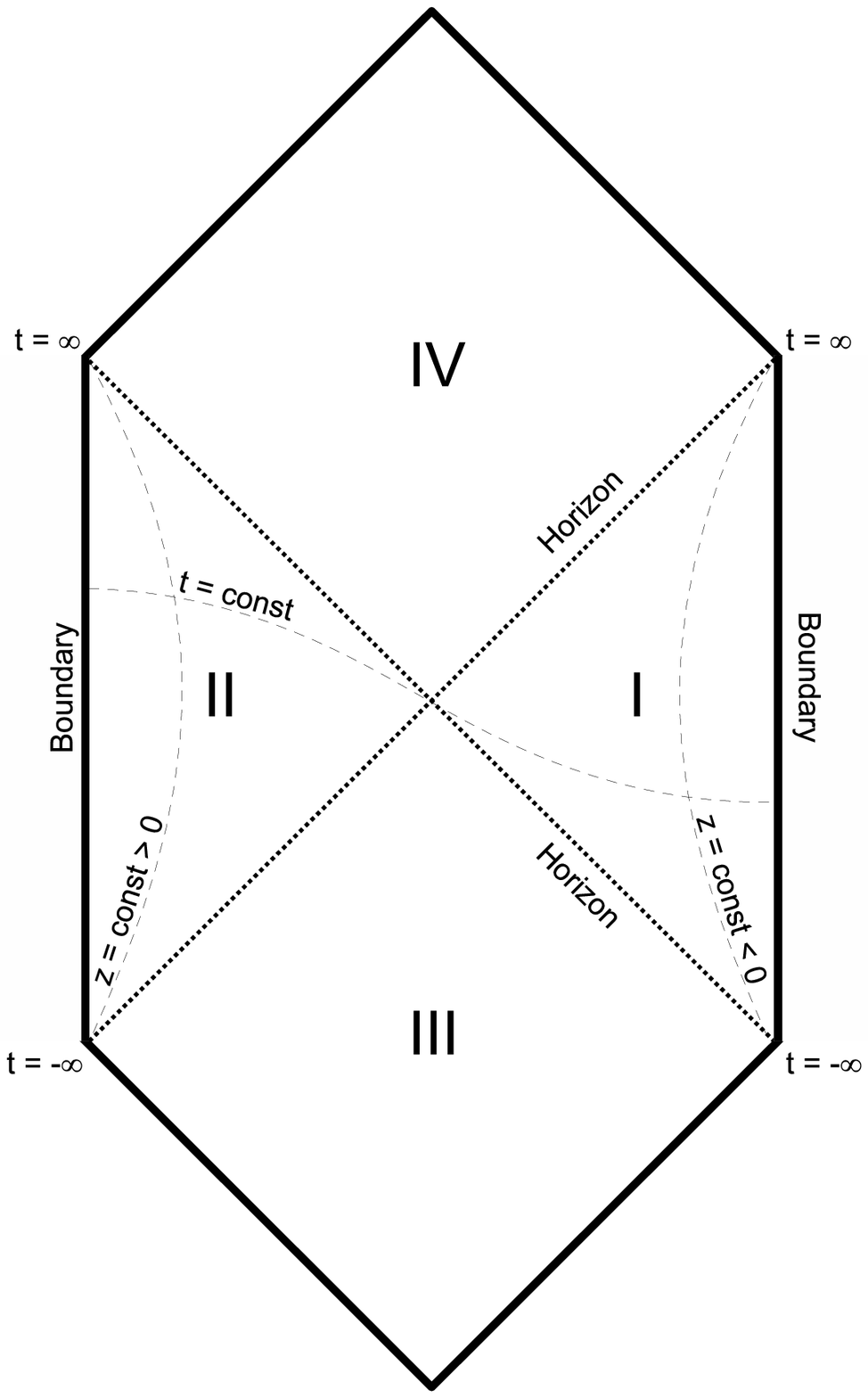}
\vskip-2.2cm
\caption[FIG:A]{Left: plot of the light cones corresponding to the solutions with $\Lambda<0$. Right: Penrose diagram of the same solution. Region I (II) corresponds to $z>0$ ($z<0$), while regions III and IV are absent on the original solution. Since the changes of coordinates needed to strip the conformal factor mix time with space, the Penrose diagram must be considered only as an approximate description, valid only at low energies.}
\label{fig:penrose1}
\end{figure}

To investigate the presence of a horizon at $z=z_o$ where the lapse function vanishes, again we evaluate the time taken by a light signal emitted for an observer that sits at an arbitrary point $z=z_1$ to reach a different point $z=z_2$. It is now given by $\Delta\tau(z_2,z_1)=e^{\frac{V_\pm(z_1)}2}(t(z_2)-t(z_1))$, with
\ba
t(z)&=& \int^{z}\!\!\!dz \;e^{-\frac{V_\pm(z)}2} = \n
&=&
\frac{(-1)^{q_\pm(\lambda)}p_\pm(\lambda)}{\sqrt{-\Lambda_W}}
\ B_{\!-1/\xi(z)}\!\left[1\!+\!q_\pm(\lambda),-q_\pm(\lambda)\right]
\ea
where we used a generalization of our previous shorthand notation $\xi(z)=\sinh^2\left({\sqrt{\Lambda_W}}(z-z_o)/p_\pm(\lambda)\right)$. The number $q_\pm(\lambda)$ has been defined as
\be
q_\pm(\lambda)=
{\frac{p_\pm(\lambda )-2\lambda}{3p_\pm(\lambda)-2}}
\ee
By choosing $z=z_o$, we have $\xi(z)=0$ and the Euler incomplete Beta function $B_{\!-1/\xi(z)}[1\!+\!q_\pm(\lambda),-q_\pm(\lambda)]\sim (-1)^{q_\pm(\lambda)}\log[\xi(z)]$ diverges. This implies that $\Delta \tau(z_o,z_1)$ is infinite, which allows us to identify the center of the geometry $z=z_o$ as a horizon. The plot of the light cones corresponding to the present solution is given at Fig. \ref{fig:penrose1}, where its approximate Penrose diagram was also included.
   
~
  
Notice that these solutions are analytic in the parameter $\lambda$ in all its range except at $\lambda=1$. There, the solution with the $+$ sign is analytic and approaches the solution for $\lambda=1$ given in (\ref{lambda1}). On the other hand, the solution with the $-$ sign is non-analytic at that point, since $p_-(\lambda)$ vanishes there.
\subsubsection{Degenerate solutions}
\label{degenerate}
In the present $\Lambda_W<0$ case, eqs.(\ref{ecuaN})-(\ref{ecua3}) allow for an additional (degenerate) branch of solutions for any value of $\lambda$. It is given by
\ba&&
e^{U_d(z)}=e^{2\sqrt{-\Lambda_W}\,(z-z_o)}\,,\n&&
e^{V_d(z)}=\mbox{arbitrary function}\,.
\ea
The existence of such an infinite branch of solutions, for which the lapse function is not determined, was already pointed out in the spherically symmetric case in \cite{Lu:2009em} 
and in the warped BTZ string context in \cite{Cho:2009fc}.
As shown in \cite{Lu:2009em}, this degeneracy is closely related to the detailed balance condition, and is lifted by an small violation of it.

\subsection{Solutions with $\Lambda_W>0$}
\label{Lambda_pos}
\subsubsection{Solutions with $\lambda=1$}
\label{Lambda_pos_lambda_1}
Proceding as in the previous sections, we first assume that $\lambda=1$. The corresponding solution reads
\ba&&
e^{U_o(z)}= \cos\left(\frac{3\sqrt{\Lambda_W }}{4}  (z-z_o)\right)^{8/3}\,,
\n&&
e^{V_o(z)}=\cos\left(\frac{3\sqrt{\Lambda_W }}{4} (z-z_o) \right)^{2/3} \sin\left(\frac{3\sqrt{\Lambda_W }}{4} (z-z_o)\right)^2\,,
\ea
where, as before, $z_o$ is a constant of integration, the solution being ${\mathbb Z}_2$ symmetric around $z=z_o$. The lapse function is real in the region $|z-z_o|\leq 2\pi/3\sqrt\Lambda_W$ and imaginary otherwise. Then the surfaces $z=z_o\pm2\pi/3\sqrt\Lambda_W$ define the boundaries of spacetime. At those boundaries, the lapse function and the function $e^{U_o(z)}$ vanish.

To have a physical interpretation of the solution, we evaluate the curvature scalar, that takes the form
\be
R^{(4)}= \frac{3}{4}\Lambda_W\left(11-5 \tan\left(\frac{3}{4} \sqrt{\Lambda_W} (z-z_o)\right)^2\right)\underset{{z\to z_o\pm \frac{2\pi}{3\sqrt{\Lambda_W}}}}\longrightarrow-\infty\,.
\ee
It takes a finite positive value proportional to $\Lambda_W$ at $z=z_o$, and it blows up to negative values at the boundaries. From the point of view of a low energy observer, that interprets the solution in the light of Einstein theory, some kind of negative energy density has to be localized close to each boundary, in order to cancel the contribution of the positive cosmological constant, resulting into a negative spacetime curvature. From our high energy point of view on the other hand, we know that we are in the presence of a purely gravitational soliton. 
Since $e^{U_o(z)}$ vanishes at the boundaries, the spatial volume of the slices  $z=z_o\pm2\pi/3\sqrt\Lambda_W$ is zero, implying that the singularities are point-like. On the other hand, all the remaining slices $z\neq z_o\pm2\pi/3\sqrt\Lambda_W$ have a finite non-zero volume  proportional to ${V}_{xy}$. Then the interpretation of this solution is that of two point like singularities sitting at the poles of the geometry $z=z_o\pm2\pi/3\sqrt\Lambda_W$ and separated by a finite physical distance $4\pi/3\sqrt\Lambda_W$. 
The space curvature confirm such view
\be
R^{(3)}= -3\Lambda_W\left(-1+\tan\left(\frac{3}{4} \sqrt{\Lambda_W} (z-z_o)\right)^2\right)\underset{{z\to z_o\pm \frac{2\pi}{3\sqrt{\Lambda_W}}}}\longrightarrow-\infty\,,
\ee
being again singular at the boundaries, while at the center it is positive and finite, proportional to $\Lambda_W$.

\begin{figure}[h]
\setlength\unitlength{1mm}
\includegraphics[height=4.5cm]{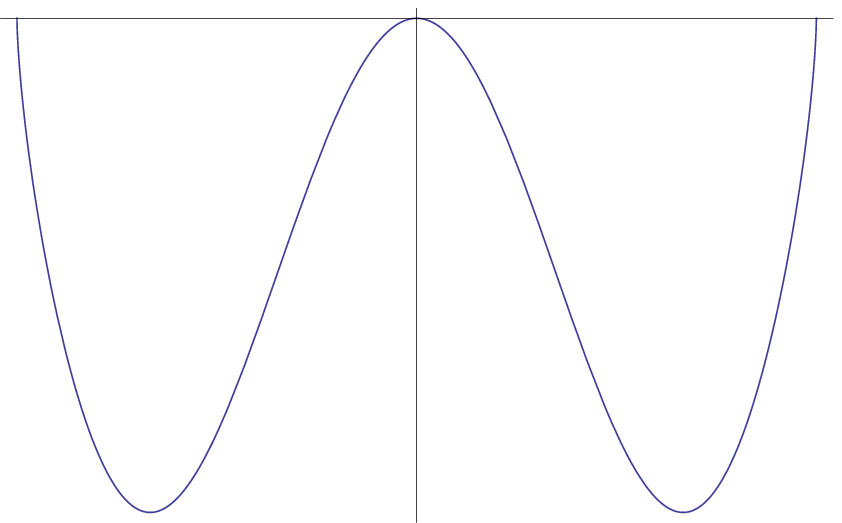}
\put(-41,48){$-e^{V_+(z)}$}
\put(2,44){$z-z_o$}\,\,\,\,\,\,\,\,\,\,\,\,
\includegraphics[height=4.5cm]{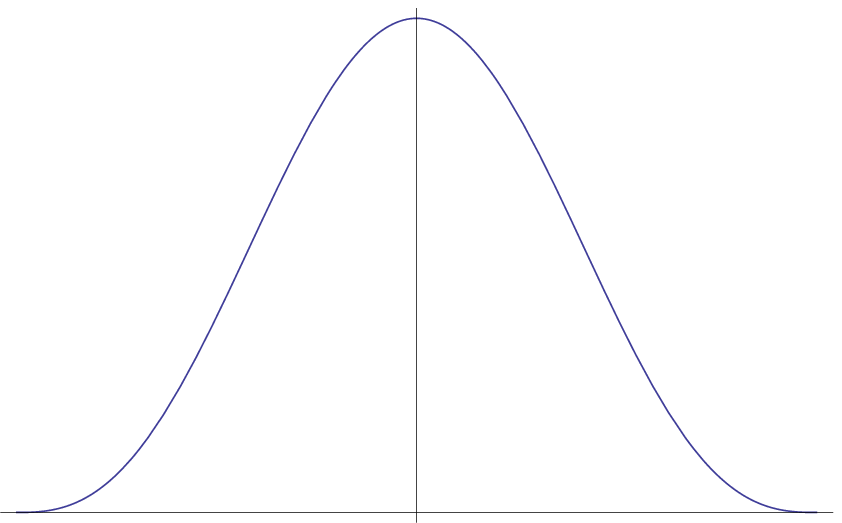}
\put(-41,46){$e^{U_+(z)}$}
\put(2,3){$z-z_o$}\\
~\\~\\
\setlength\unitlength{1mm}
\includegraphics[height=4.5cm]{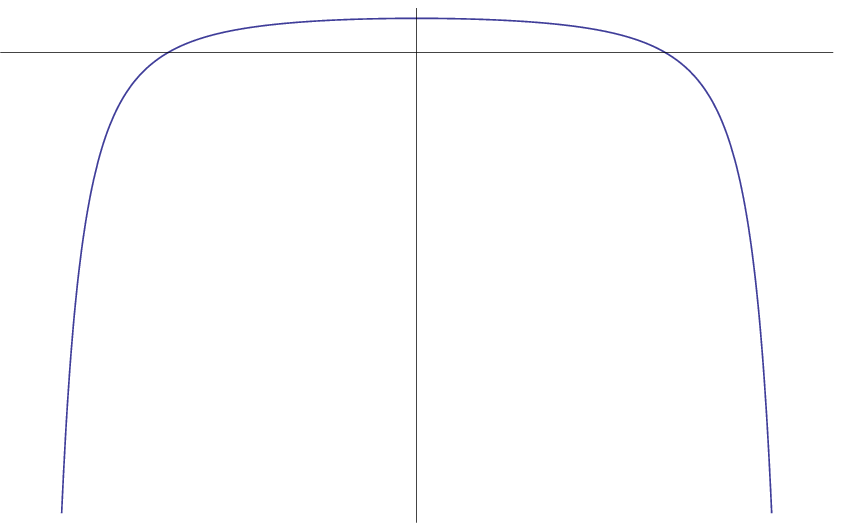}
\put(-41,46){$R^{(4)}(z)$}
\put(1,41){$z-z_o$}\,\,\,\,\,\,\,\,\,\,\,\,\,\,\,\,\,
\includegraphics[height=4.5cm]{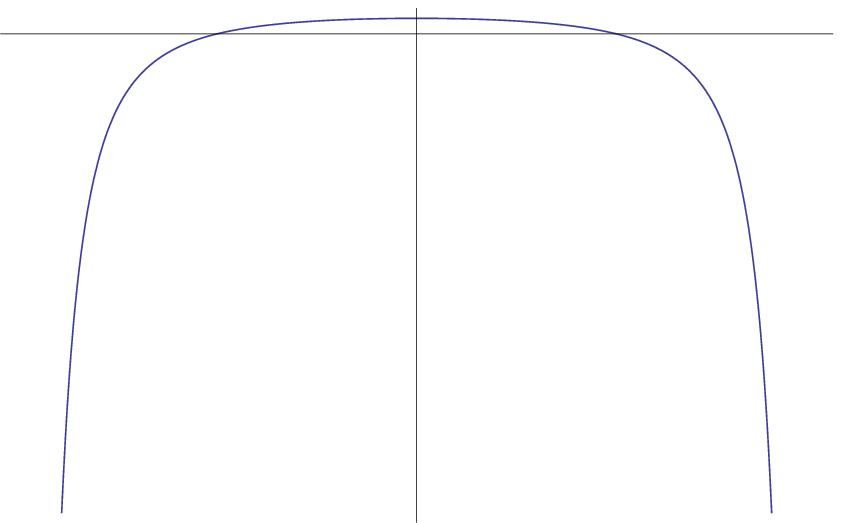}
\put(-41,46){$R^{(3)}(z)$}
\put(2,43){$z-z_o$}
\caption[FIG:A]{Plots of the lapse function $g_{00}=-e^{V_o(z)}$ (up-left) and the spatial volume function $g_{xx}=g_{yy}=e^{U_o(z)}$ (up-right), spacetime curvature $R^{(4)}(z)$ (down-left) and space curvature $R^{(3)}(z)$ (down-right) for  $\Lambda_W>0$ and $\lambda=1$}.
\label{fig.LambdaPositivo.lambda1}
\end{figure}

Plots of the solutions together with their spatial and spacetime curvatures can be seen in Fig.\ref{fig.LambdaPositivo.lambda1}. 

~

To investigate whether the singularity we just found is hidden behind a horizon, we proceed as in the previous section. 
The time taken by a light signal emitted at $z=z_1$ to reach $z=z_2$ is $\Delta\tau(z_2,z_1) = e^{\frac{V_o(z_1)}2}(t(z_2)-t(z_1))$ where now
\ba
t(z)
&=& 
-\int^{z}\!\!\!dz \;e^{-\frac{V_o(z)}2} = 
\n
&=&
\frac{-1}{3\sqrt{\Lambda_W}}\, \left(2 \sqrt{3} \cot ^{-1}\!\!\left(\frac{2 \chi ^{{1}/{3}}(z)+1}{\sqrt{3}}\right)+\ln \left(1-\frac{3 \chi^{{1}/{3}}(z)}{\chi ^{{2}/{3}}(z)+\chi^{{1}/{3}}(z)+1}\right)\right) 
\ea
where we used the notation $\chi(z)=\cos^2\left(3{\sqrt{\Lambda_W}}(z-z_o)/4\right)$. By choosing $z=z_o$, we have $\chi(z)=1$ and the logarithm diverges. This implies that the slice at the center of the geometry $z=z_o$ is a horizon. 
On the other hand, by choosing $z=z_o\pm2\pi/3\sqrt\Lambda_W$ we have $\chi(z)=0$, and the integral is finite. This implies that there are no horizons sitting at the singularities. In conclusion, from the point of view of any observer, one of the singularities is beyond a horizon and the other is naked. 
We included the corresponding plot of the light cones in Fig.\ref{fig:horizons} and the approximate Penrose diagram in Fig.\ref{fig:penrose2}.
\subsubsection{Solutions with $\lambda\neq1$}
\label{Lambda_pos_lambda_neq_1}
Next, we move to the case $\lambda\neq1$ . Then, 
solving eq.(\ref{ecuaN}) for $U(z)$ and replacing the solution into eq.(\ref{ecua3}) to get $V(z)$, we have
\ba&&
e^{U_\pm(z)}=
\cos\left(
\frac{\sqrt{\Lambda_W} }{p_\pm(\lambda)}\,(z-z_o)
\right)^{2p_\pm(\lambda)}\,,
\n&&
e^{V_\pm(z)} = \left({\cos\left(\frac{\sqrt{\Lambda_W} }{p_\pm(\lambda)}(z-z_o)\right)^{\frac{5 p_\pm(\lambda )-4\lambda-2}{3p_\pm(\lambda)-2}} \sin\left(\frac{\sqrt{\Lambda_W}}{p_\pm(\lambda )}(z-z_o)\right)}\right)^2\,,
\ea
where, as before, $z_o$ is a constant of integration, and $p_\pm(\lambda)$ is given by expression (\ref{ppm}). Again the solution does not exists for $\lambda<1/3$, while for $\lambda>1/3$ we get two branches of solutions identified with the $\pm$ subindices according to the choice of sign in the $p_\pm(\lambda)$.
\begin{figure}[h]
\setlength\unitlength{1mm}
\includegraphics[height=4.5cm]{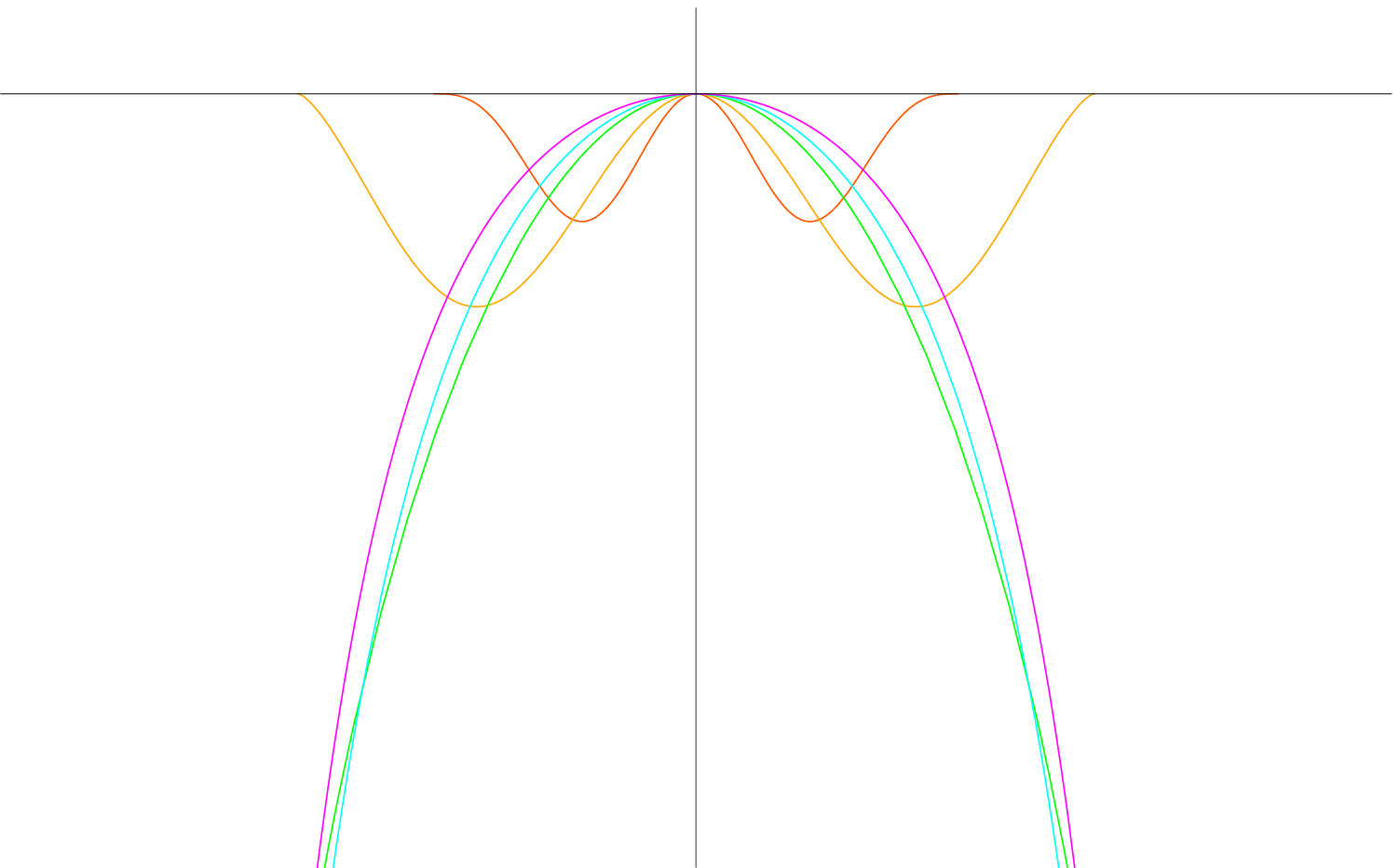}
\put(-41,46){$-e^{V_+(z)}$}
\put(2,41){$z-z_o$}\,\,\,\,\,\,\,\,\,\,\,\,
\includegraphics[height=4.5cm]{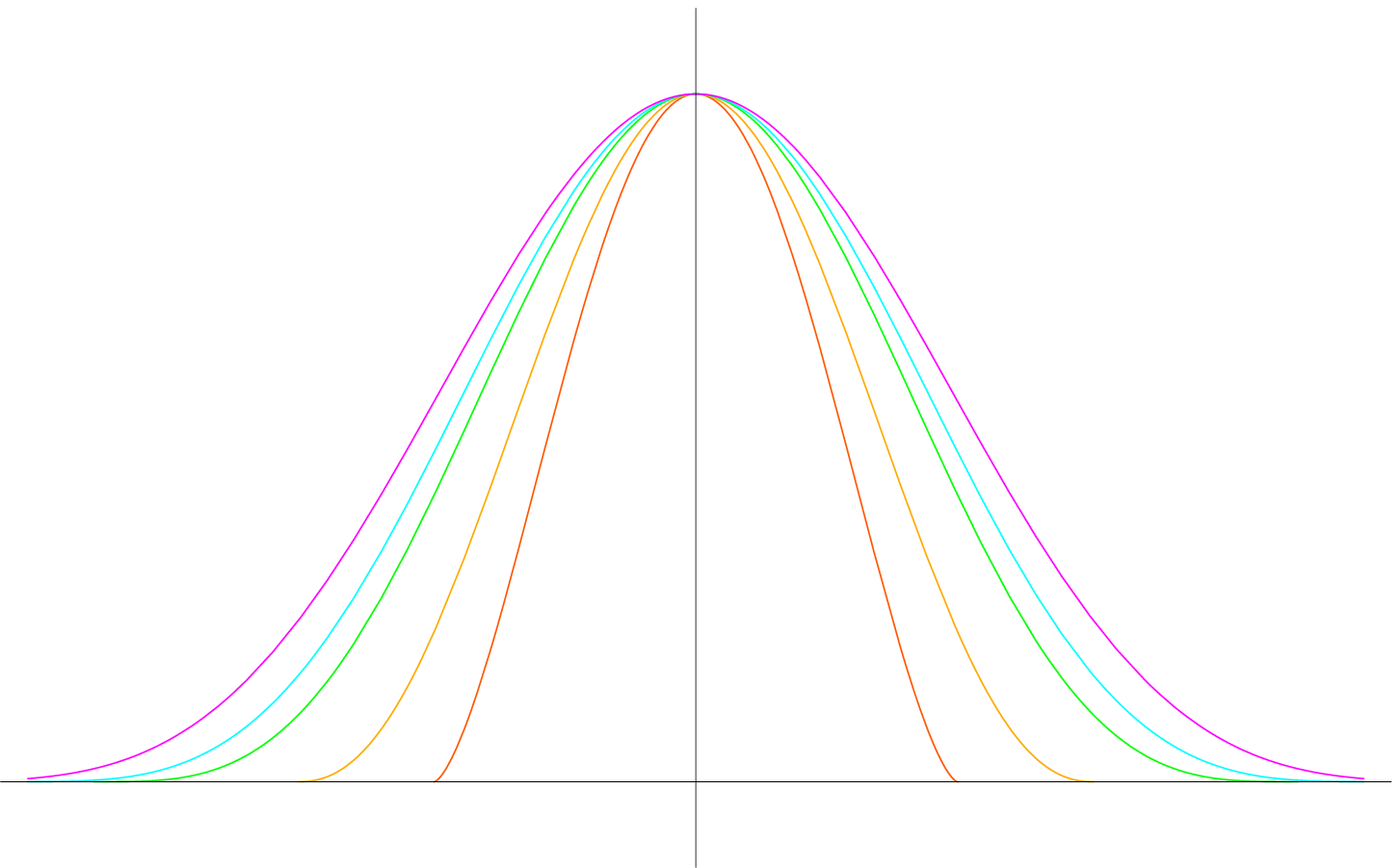}
\put(-41,46){$e^{U_+(z)}$}
\put(2,5){$z-z_o$}\\
~\\~\\
\setlength\unitlength{1mm}
\includegraphics[height=4.5cm]{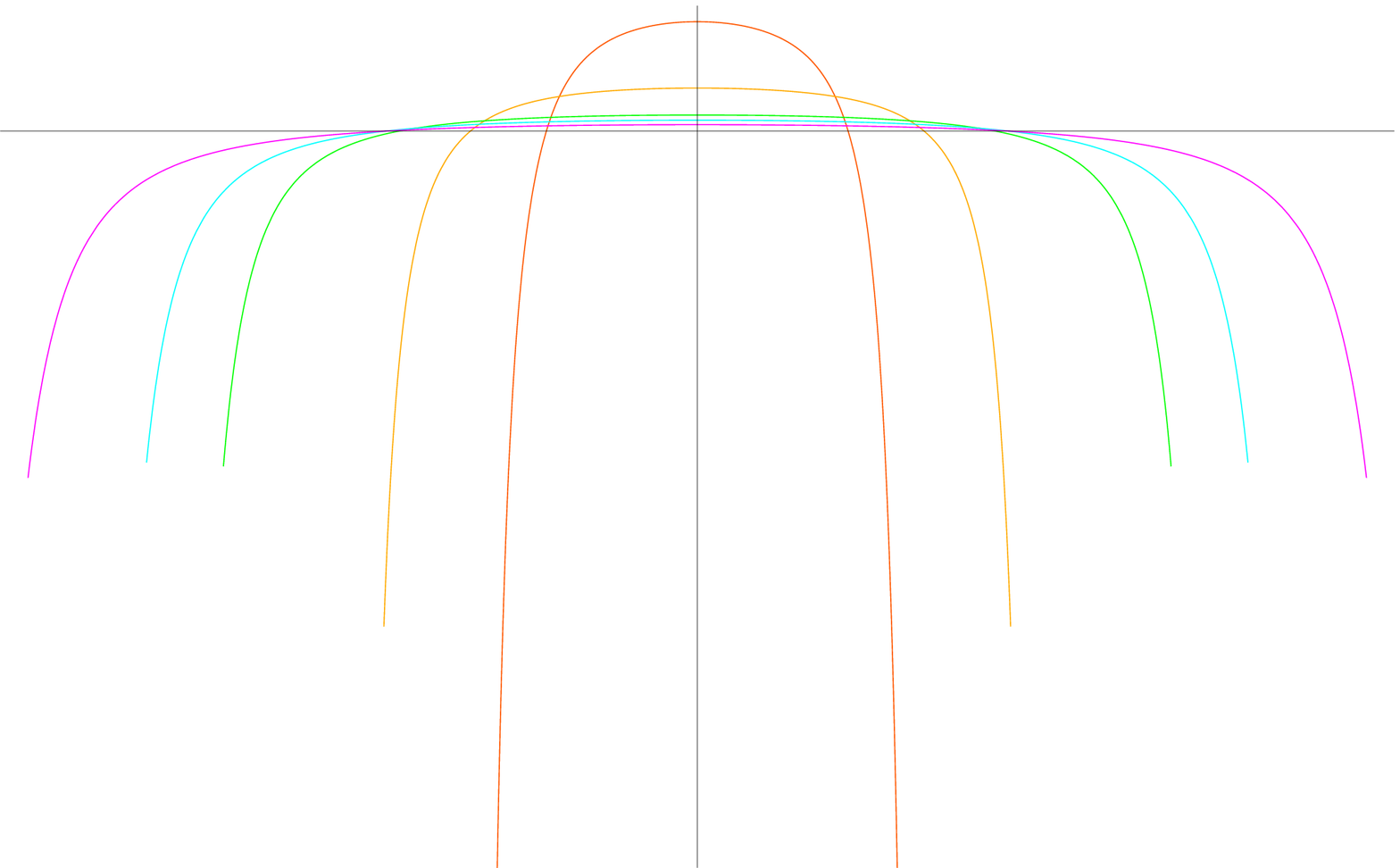}
\put(-41,46){$R^{(4)}(z)$}
\put(2,39){$z-z_o$}\,\,\,\,\,\,\,\,\,\,\,\,\,\,
\includegraphics[height=4.5cm]{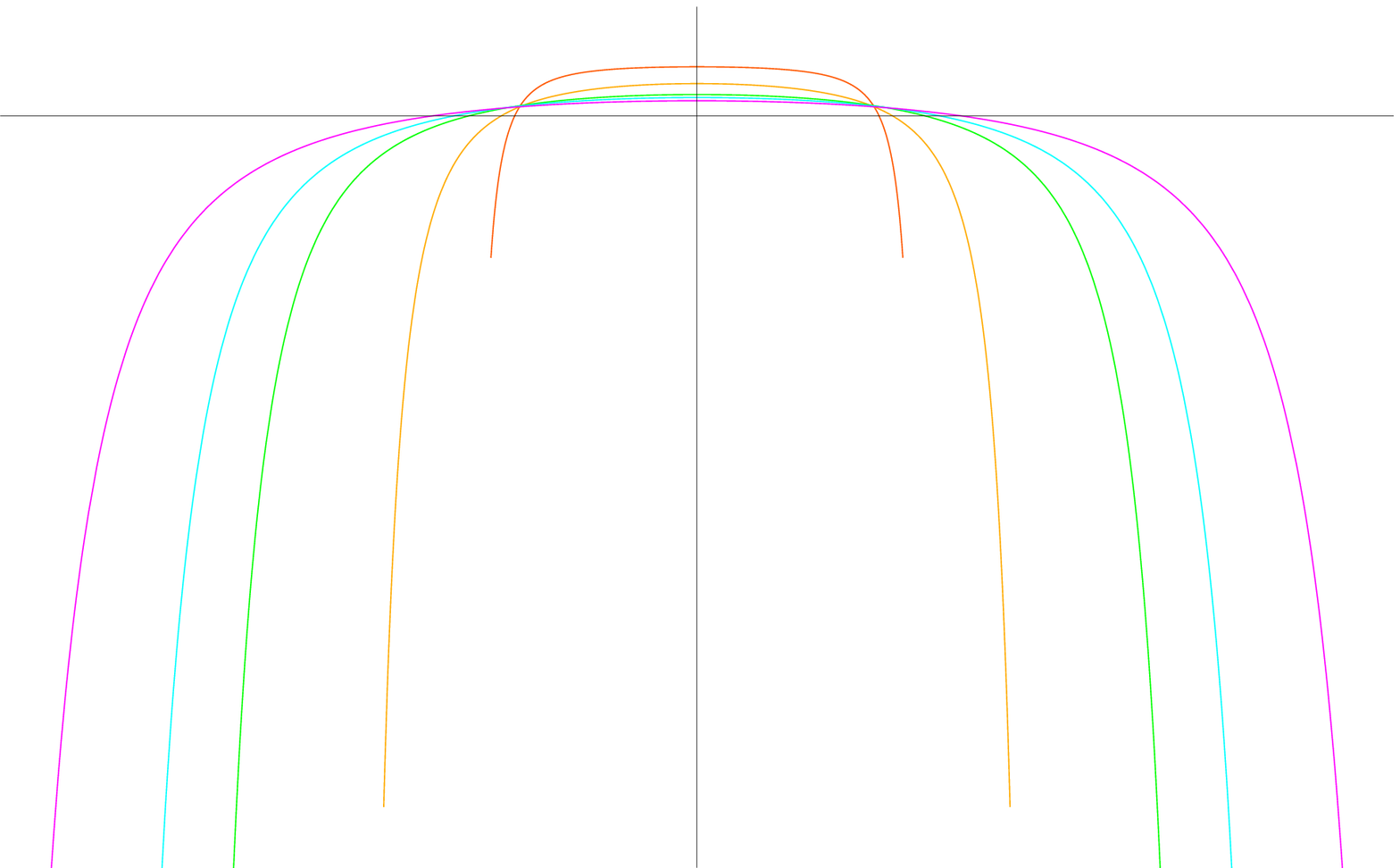}
\put(-41,46){$R^{(3)}(z)$}
\put(2,40){$z-z_o$}
\caption[FIG:A]{Plots of the lapse function $g_{00}=-e^{V_+(z)}$ (up-left) and the spatial volume $g_{xx}=g_{yy}=e^{U_+(z)}$ (up-right), spacetime curvature $R^{(4)}(z)$ (down-left) and space curvature $R^{(3)}(z)$ (down-right) for  $\Lambda_W>0$,  for different values of $\lambda$. Notice the change of behavior of the lapse function at the horizons at $\lambda=3$, in the top left figure.}
\label{fig.LambdaPositivo.++}
\end{figure}
\begin{figure}[h]
\setlength\unitlength{1mm}
\includegraphics[height=4.5cm]{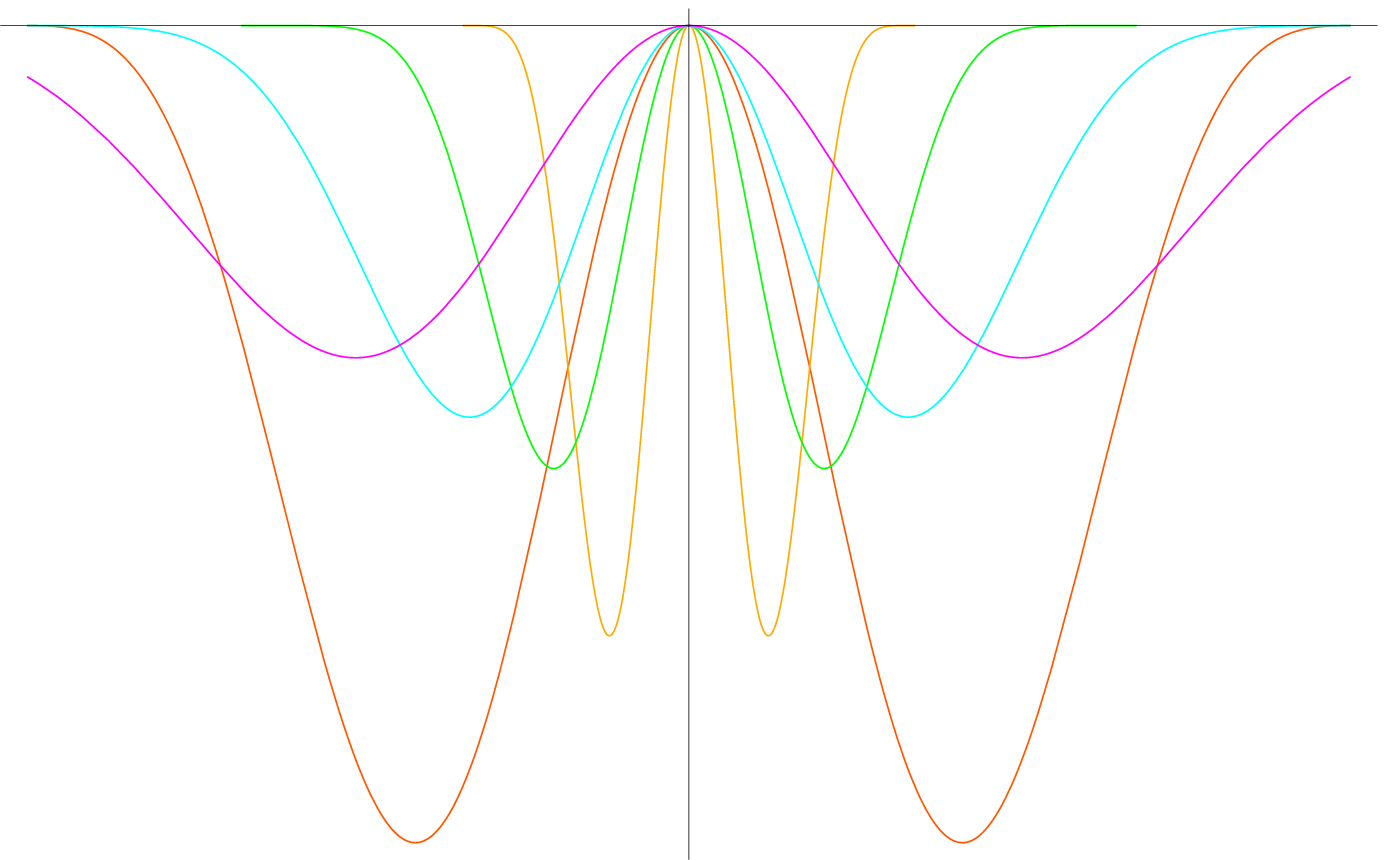}
\put(-41,46){$-e^{V_-(z)}$}
\put(2,44){$z-z_o$}\,\,\,\,\,\,\,\,\,\,\,\,
\includegraphics[height=4.5cm]{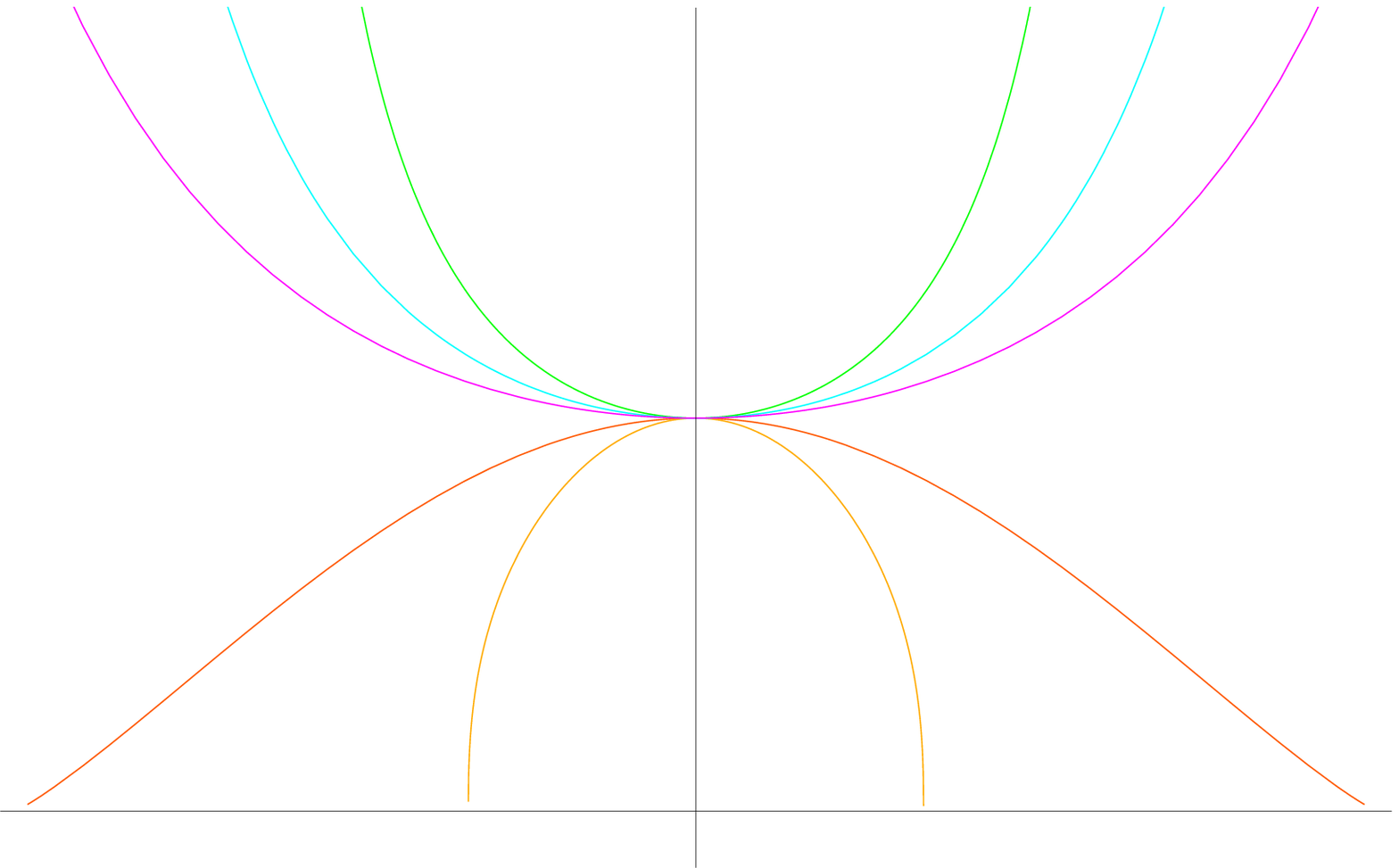}
\put(-41,46){$e^{U_-(z)}$}
\put(2,3){$z-z_o$}\\
~\\
~\\
\setlength\unitlength{1mm}
\includegraphics[height=4.5cm]{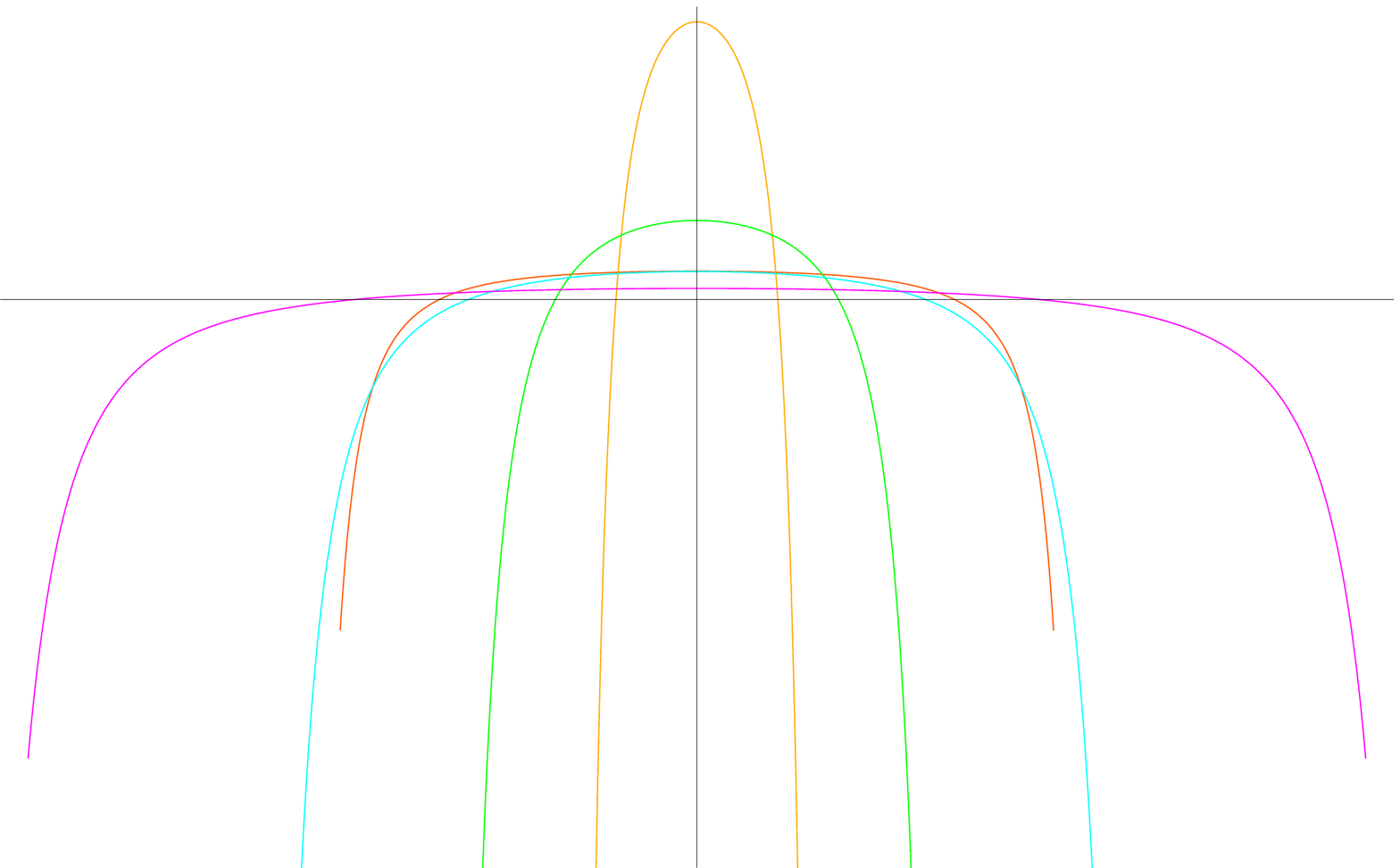}
\put(-41,46){$R^{(4)}(z)$}
\put(2,30){$z-z_o$}\,\,\,\,\,\,\,\,\,\,\,\,
\includegraphics[height=4.5cm]{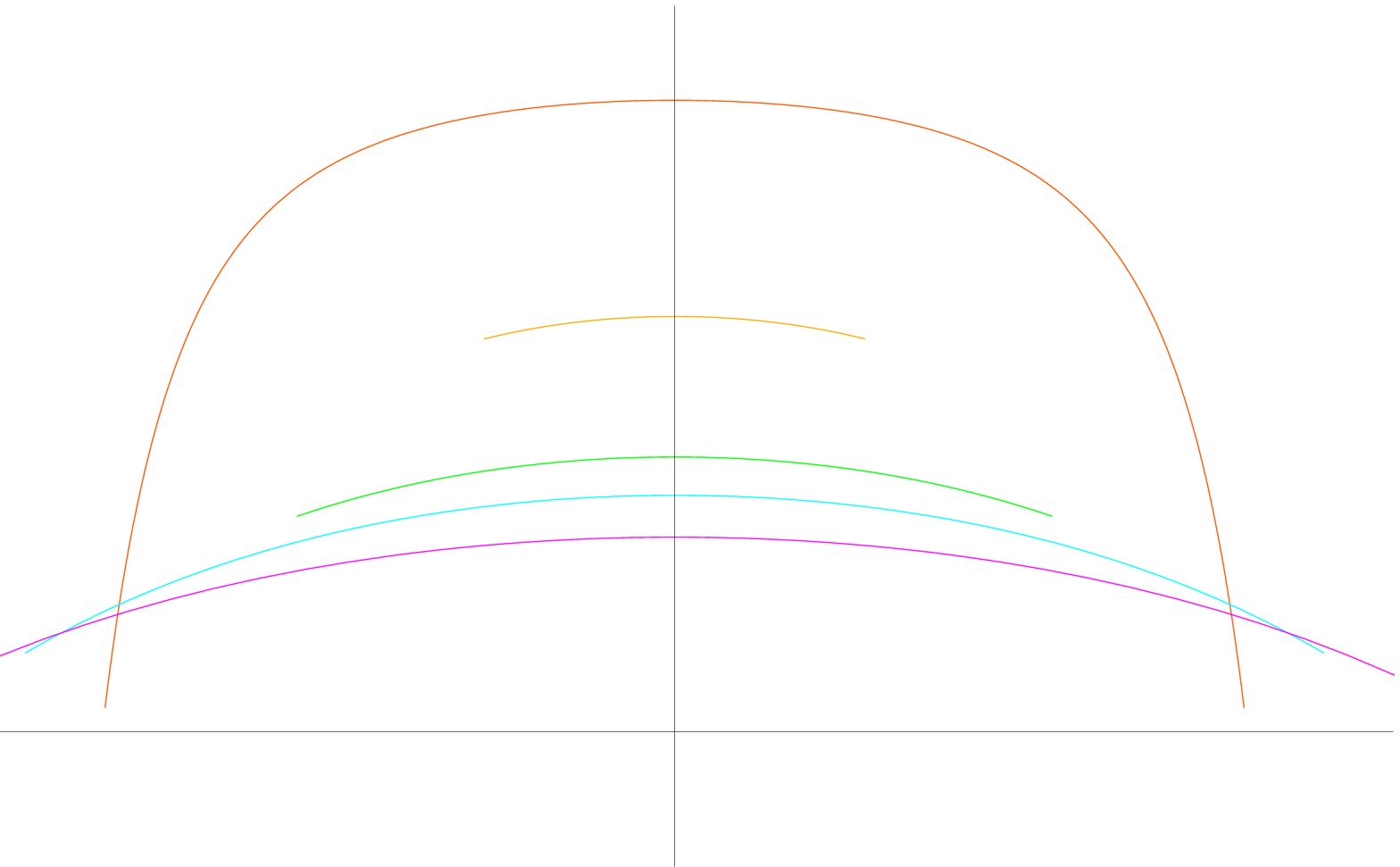}
\put(-42,46){$R^{(3)}(z)$}
\put(2,8){$z-z_o$}
\caption[FIG:A]{Plots of the lapse function $g_{00}=-e^{V_-(z)}$ (up-left) and the spatial volume function $g_{xx}=g_{yy}=e^{U_-(z)}$ (up-right), spacetime curvature $R^{(4)}(z)$ (down-left) and space curvature $R^{(3)}(z)$ (down-right) for  $\Lambda_W>0$, for different values of $\lambda$. Notice the change of behavior of the lapse function at the horizons at $\lambda=1$, in the top right figure.}.
\label{fig.LambdaPositivo.+-}
\end{figure}

For both branches, the lapse function is real in the region $|z-z_o|<\pi p_\pm(\lambda)/2\sqrt{\Lambda_W}$ and imaginary otherwise. Then the surfaces $z=z_o\pm\pi p_\pm(\lambda)/2\sqrt{\Lambda_W}$ define the boundaries of space. At the boundary, the behavior of the solution depends on the value of $\lambda$. For the solution with the $+$ sign, the exponent of the cosine in the lapse function changes sign at $\lambda=3$, implying that for $\lambda<3$ the lapse function vanishes at the boundary, while for $\lambda>3$ it diverges. On the other hand, for the solution with the minus sign, the exponent of the cosine in the spatial volume function changes sign when $\lambda=1$, resulting in a solution whose spatial volume vanishes at the boundary for $\lambda<1$ and diverges for $\lambda>1$. 

The spacetime curvature for the above solutions read
\be
R^{(4)}=\tilde R^{(4)}(\lambda)+(R^{(4)}_{z_o}(\lambda)-\tilde R^{(4)}(\lambda))\, \text{sec}\left(\frac{\sqrt{\Lambda_W} (z-z_o)}{p_\pm(\lambda)}\right)^2\,,
\ee
where
\be
\tilde R^{(4)}(\lambda)=\frac{-2\Lambda_W\left(-16 (1+\lambda )^2+p_\pm(\lambda)(48 (1+\lambda )-p_\pm(\lambda) (20-24 \lambda +3 p_\pm(\lambda)(4+9 p_\pm(\lambda))))\right)}{(2-3 p_\pm(\lambda))^2 p_\pm(\lambda)^2}\,,
\ee
and
\be
R^{(4)}_{z_o}(\lambda)=\tilde R^{(4)}(\lambda)-
\frac{2\Lambda_W\left(8 \lambda  (1+2 \lambda )+p_\pm(\lambda)(-4 (1+3 \lambda )+p_\pm(\lambda) (14-24 \lambda +3 p_\pm(\lambda) (-8+9 p_\pm(\lambda))))
\right)}{(2-3 p_\pm(\lambda))^2 p_\pm(\lambda)^2}\,.
\ee
Again it is clear that the solution is singular at the boundaries, where the function $\sec\left(\sqrt{-\Lambda_W} (z-z_o)/p_\pm(\lambda)\right)$ diverges. These singularities are separated by a finite distance $\pi p_\pm(\lambda)/\sqrt{\Lambda_W}$. 

Since the spatial volume function vanishes at the boundaries whenever $p_\pm(\lambda)$ is positive, in such case the singularities will be point-like. This happens for any $\lambda$ in the $+$ branch of solutions, and for $\lambda<1$ in the $-$ branch. In the rest of the $-$ branch, that is for $\lambda>1$, the volume of the boundaries diverges, implying that the singularities can be interpreted as two parallel purely gravitational membranes, sitting at the boundaries of spacetime. 

Plots of the solutions with the $+$ sign with the corresponding spacetime and space curvatures for different values of $\lambda$ can be seen in Fig.\ref{fig.LambdaPositivo.++}. On the other hand, plots of the solutions with the $-$ sign with the corresponding spacetime and space curvatures for different values of $\lambda$ can be seen in Fig.\ref{fig.LambdaPositivo.+-}.

The time taken by a light signal emitted from an observer sitting at $z=z_1$ to reach a point $z=z_2$ is given by $\Delta\tau(z_2,z_1)= e^{\frac{V_\pm(z_1)}2}(t(z_2)-t(z_1))$ with
\ba
t(z)&=&\int^{z}\!\!\!dz \;e^{-\frac{V_\pm (z)}2} =
\n
&=&
-\frac{p_\pm(\lambda)}{2\sqrt{\Lambda_W}q_\pm(\lambda )}\ {\chi^{q_\pm(\lambda )}(z)}\!~_2F_1\!\left[1, q_\pm(\lambda ); 1+q_\pm(\lambda ); \chi(z)\right]
\ea
where the notation is now $\chi(z)=\cos^2\left({\sqrt{\Lambda_W}}(z-z_o)/p_\pm(\lambda)\right)$, the symbol $_2F_1$ representing the hypergeometric function. Since such function diverges at $\chi(z)=1$, that is at $z=z_o$, the time taken by a light signal to reach such slice is infinite. In consequence, as in the case $\lambda=1$, there will be a horizon at the center of space. On the other hand, the prefactor $\chi^{q_\pm(\lambda)}(z)$ vanishes at $z=z_o\pm\pi p_\pm(\lambda)/2\sqrt{\Lambda_W}$ whenever the exponent $q_\pm(\lambda)$ is positive, and diverges whenever it is negative. 
This implies that, as in the case $\lambda=1$, there will be no horizons at the singularities for $\lambda>1/2$ in the $+$ branch. For the rest of the $+$ branch, that is for $\lambda<1/2$, as well as for all values of $\lambda$ in the $-$ branch, there will be a horizon sitting at each of the singularities. An observer at any point of the bulk will regard one or both of the singularities as hidden behind horizons according to the value of $\lambda$ and to the branch he/she is probing. 

\begin{figure}[h]
\setlength\unitlength{1mm}
\includegraphics[height=10cm]{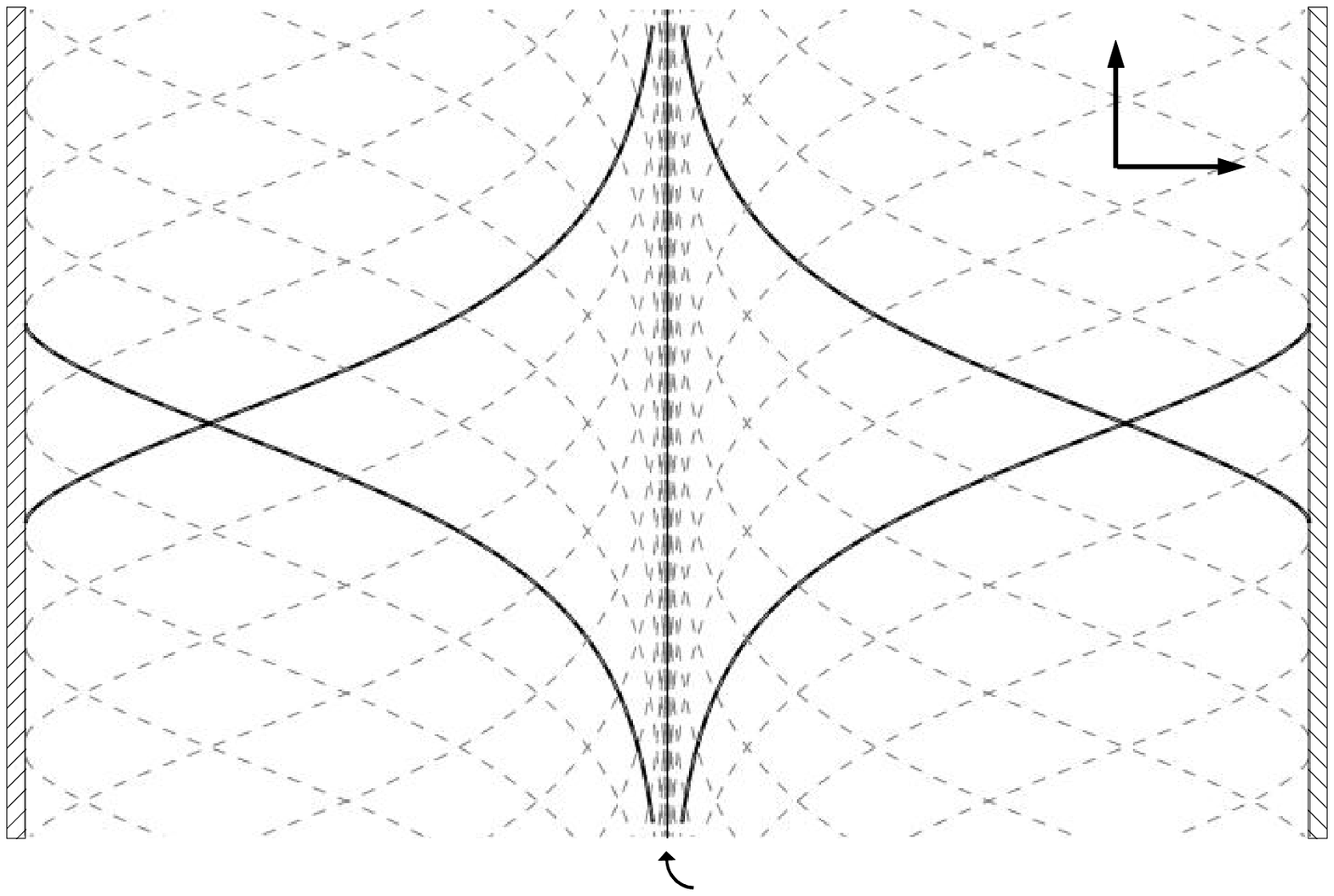}
\put(-33,55){Horizon}
\put(-17,92){$t$}
\put(-10,85){$z$}
\put(-4.5,69){{\begin{sideways}Singularity\end{sideways}}}
\put(-67.2,69){{\begin{sideways}Singularity\end{sideways}}}
\,\,\,\,\,\,\,\,\,\,\,\,
\,\,\,\,\,\,\,\,\,\,\,\,
\includegraphics[height=10cm]{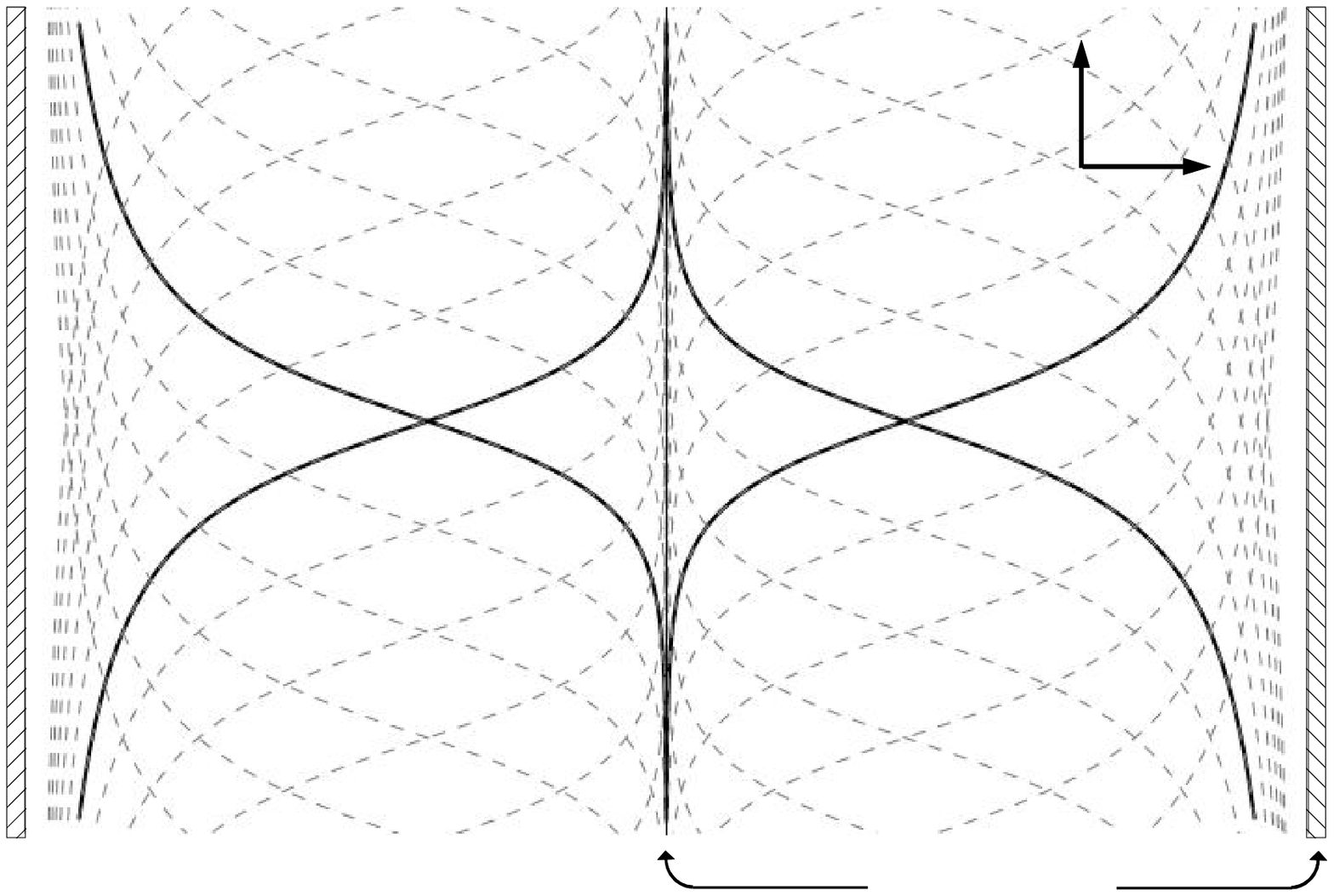}
\put(-25.7,55){Horizon}
\put(-18,92){$t$}
\put(-12,85){$z$}
\put(-4.5,69){{\begin{sideways}Singularity\end{sideways}}}
\put(-67.2,69){{\begin{sideways}Singularity\end{sideways}}}
\vskip-5.3cm
\caption[FIG:A]{Plots of the light cones for the $\Lambda_W>0$ solutions. On the left we see the plot corresponding to the case in which there is a single horizon at the center of space, that is $\lambda>1/2$ in the $+$ branch, which includes the nondegenerate solution with $\lambda=1$. On the right, the case in which the singularities are hidden behind horizons, that is $\lambda<1/2$ in the $+$ branch and all values of $\lambda$ in the $-$ branch.}
\label{fig:horizons}
\end{figure}

\begin{figure}[h]
\vskip-1.5cm
\includegraphics[height=12cm]{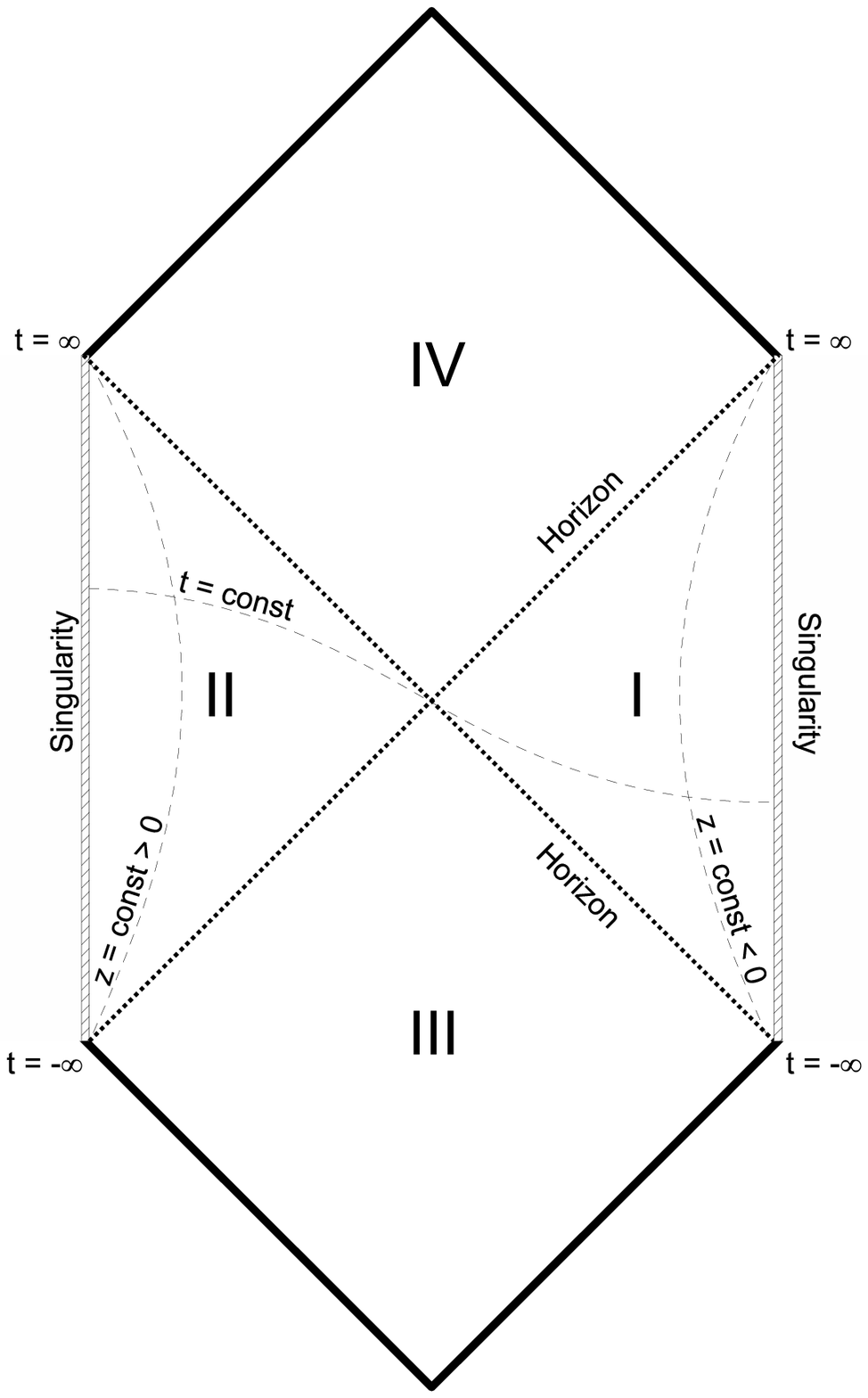}
\includegraphics[height=12cm]{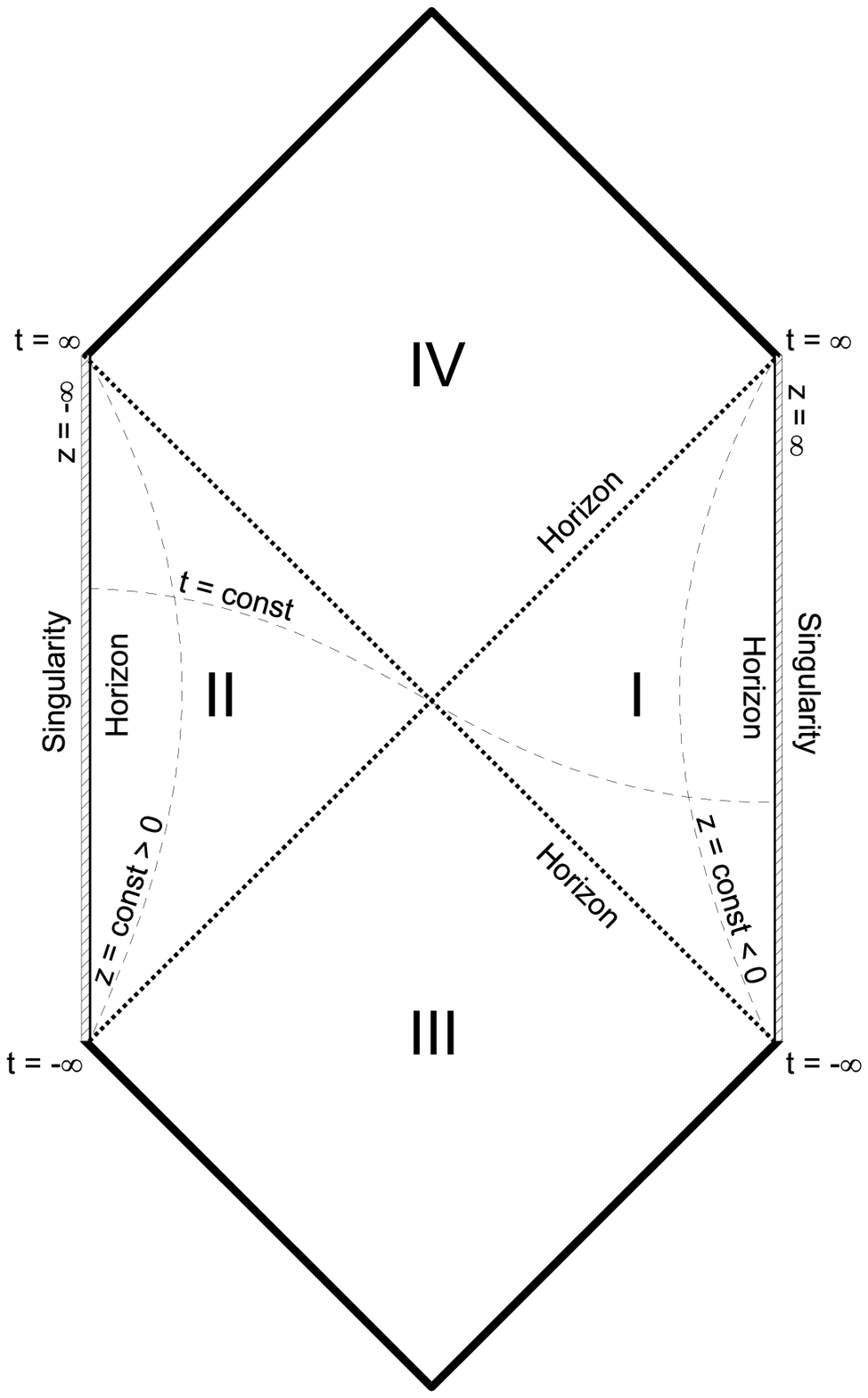}
\vskip-2cm
\caption[FIG:A]{Penrose diagrams of the $\Lambda>0$ solutions. The left (right) figure correspond to $\lambda>1/2$ in the $+$ branch ($\lambda<1/2$ in the $+$ branch and all values of $\lambda$ in the $-$ branch). Regions I (II) correspond to $z>0$ ($z<0$), while regions III and IV are absent in the original solution. In the left plot the singularities are naked, while in the right they are hidden behind horizons. Again, it must be kept in mind that these diagrams are only valid as a low energy approximation, since in constructing them we used changes of coordinates that mix time with space.}
\label{fig:penrose2}
\end{figure}

As conclusion, the solutions of the $+$ branch will consist in two point-like singularities separated by a finite distance, which are hidden behind horizons whenever $\lambda<1/2$. On the other hand, the solutions of the $-$ branch will consist for $\lambda<1$  in two point like singularities, and for $\lambda>1$ in two parallel membranes, in both cases the singularities being hidden behind horizons.
Plots of the light cones in both cases can be seen in Fig.\ref{fig:horizons}, the approximate Penrose diagrams can be seen in Fig.\ref{fig:penrose2}.
\section{Discussion}

We explored membrane solutions to Ho\v rava nonrelativistic theory of gravity when the detailed balance condition is satisfied. We found that for arbitrary values of the parameter $\lambda>1/3$, branches of membrane solutions exist. 

For the particular case $\lambda=1$ there is a single branch that corresponds to a ${\mathbb Z}_2$ symmetric spacetime. In the case of negative cosmological constant, the spacetime and spatial curvatures have a peak at $z=z_o$, that allows us to identify the solution as a membrane sitting at the center of space. There is a horizon at the membrane location. The curvatures become proportional to $\Lambda_W$ at large distances, where the metric correspond to an AdS spacetime. On the other hand, in the case of positive cosmological constant, the space has boundaries at $|z-z_o|=2\pi/3\sqrt{\Lambda_W}$ beyond which the metric becomes complex. The spacetime and space curvatures diverge at the boundaries. The spatial volume of the slices containing the singularities vanishes, which implies that the singularities are point like. There is a horizon at the center of space, implying that from the point of view of any observer, one of the singularities is hidden beyond a horizon an the other is naked.  On the other hand, both curvatures are finite and positive at the intermediate space.

For generic values of $\lambda$ in the region $1/3<\lambda\neq1$, two branches appear that correspond to ${\mathbb Z}_2$ symmetric solutions. In the case of negative cosmological constant, the space is unbounded and the space curvature asymptotes a constant value proportional to $\Lambda_W$. The same is true for the spacetime curvature, but with a proportionality factor that is a function of $\lambda$. The asymptotic metric corresponds to a Lifshitz spacetime. Again, the curvatures being finitely peaked at the center of space, we interpret the solution as representing a membrane sitting there. The  curvatures at the center depend on the value of $\lambda$. There is a horizon at the membrane. On the other hand, in the case of positive cosmological constant, the space is bounded and the curvatures are singular at the boundaries. The spatial volume of those boundaries vanishes for $\lambda<1$ in the $-$ branch and for all values of $\lambda$ in the $+$ branch, implying that the singularities are point like. On the other hand for $\lambda>1$ in the $-$ branch the area of the slices containing the singularities diverges, allowing the interpretation of them as membranes. As in the $\lambda=1$ case, there is a horizon at the center of space for any value of $\lambda$ in both branches. For $\lambda>1/2$ in $+$ branch, there is no horizon at the boundaries, while for $\lambda<1/2$ in the $+$ branch and for all values of $\lambda$ in the $-$ branch the singularities are hidden behind horizons. 

It should be kept in mind that our solutions are purely gravitational solitons, since no additional matter terms have been added to Ho\v rava action. Nevertheless, our nomenclature was inspired in the point of view of a low energy observer, according to whom the dynamics of gravity is totally covariant and given by Einstein theory. He/she would necessarily interpret our gravitational domain wall as originated on some kind of membrane-like matter sources.

For any value of $\lambda$, and additional branch exists. It is degenerate, in the sense that the lapse function is completely undetermined by the equations of motion. This behavior has been reported before in the case of spherically symmetric \cite{Lu:2009em} and warped BTZ string \cite{Cho:2009fc} solutions, and was related with the detailed balance condition \cite{Lu:2009em}. 

It is interesting to note that the solutions behave analytically in the parameter $\lambda$ for all values of $\lambda\neq 1$. The solutions with the $+$ sign approaches the regular $\lambda=1$ branch in the limit $\lambda\to1$, while the solution with the $-$ sign is not analytic in that limit. 

As possible continuations of this work, it may be interesting to study solutions where the detailed balance condition is softly broken, or where the action is extended to the so called ``healthy version'' of Ho\v rava gravity. Moreover, the study of stability under perturbations may be interesting, as well as the inclusion of a Lifshitz scalar field with a symmetry breaking potential that could provide a topologically conserved charge. 

\section{Acknowledgments}

The authors want to thank Hector Vucetich and Guillermo Silva for help and encouragement during this work. They also thank Susana Landau and Ana Mar\'\i a Platzek for helpful comments on the M.Sc. thesis that originates the present paper. This work is partially supported by ANPCyT grants PICT 00849 and 20350, and CONICET grant PIP2010-0396.

\end{document}